\documentclass[11pt]{article}
\pdfoutput=1

\usepackage{amssymb}
\usepackage{amsmath}
\usepackage{bbold}
\usepackage{color}
\usepackage{colordvi}
\usepackage{dsfont}
\usepackage{fancybox}
\usepackage[footnotesize]{caption2}
\usepackage{graphicx}
\usepackage[center,footnotesize,hang]{subfigure}
\usepackage{bbm}
\usepackage{url}
\usepackage{multirow}
\usepackage{array}
\usepackage{arydshln}
\usepackage{pifont}
\usepackage{mathrsfs}
\usepackage{fancybox}
\usepackage{cite}
\usepackage[colorlinks]{hyperref}
\usepackage{enumitem}
\usepackage{longtable}
\newcommand{\ignore}[1]{}
\usepackage{enumitem}

\newcommand{\PreserveBackslash}[1]{\let\temp=\\#1\let\\=\temp}
\newcolumntype{C}[1]{>{\PreserveBackslash\centering}p{#1}}
\newcolumntype{R}[1]{>{\PreserveBackslash\raggedleft}p{#1}}
\newcolumntype{L}[1]{>{\PreserveBackslash\raggedright}p{#1}}
\allowdisplaybreaks
\newcommand{\bq}{\begin{eqnarray}}
\newcommand{\nq}{\end{eqnarray}}

\usepackage[normalem]{ulem}

\def\bvec#1{\raise1.5ex\hbox{$\rightarrow$}\mkern-16.5mu #1}

\begin{document}
\title{
\begin{flushright}
\begin{minipage}{0.2\linewidth}
\normalsize
\end{minipage}
\end{flushright}
{\Large \bf
Leptogenesis in $SO(10)$ Models with $A_4$ Modular Symmetry
\\[2mm]}}
\date{}

\author{
Gui-Jun Ding$^{1}$\footnote{E-mail: {\tt
dinggj@ustc.edu.cn}},  \
Stephen~F.~King$^{2}$\footnote{E-mail: {\tt king@soton.ac.uk}}, \
Jun-Nan Lu$^{1}$\footnote{E-mail: {\tt
hitman@mail.ustc.edu.cn}},  \
Bu-Yao Qu$^{1}$\footnote{E-mail: {\tt
qubuyao@mail.ustc.edu.cn}}
\\*[20pt]
\centerline{
\begin{minipage}{\linewidth}
\begin{center}
$^1${\it \normalsize Department of Modern Physics, University of Science and Technology of China, \\ Hefei, Anhui 230026, China}\\[2mm]
$^2${\it \normalsize
Physics and Astronomy, University of Southampton, Southampton, SO17 1BJ, U.K.}
\end{center}
\end{minipage}}
\\[10mm]}
\maketitle
\thispagestyle{empty}

\begin{abstract}
We study the prediction for leptogenesis in two renormalizable supersymmetric $SO(10)\times A_4$ modular models in which the neutrino mass is dominantly generated by the type I seesaw mechanism. The evolution of the lepton asymmetries are described in terms of the three-flavored density matrix equations for three heavy Majorana neutrinos, where both vanishing initial condition and thermal initial condition of the right-handed neutrinos are considered. We also present an analytical approximation based on the Boltzmann equations. We find regions of parameter space compatible with the measured fermion masses and mixing parameters as well as the baryon asymmetry of the Universe. The predictions for the light neutrino masses, the effective mass in neutrinoless doble beta decay and the leptonic CP violation phases are discussed.

\end{abstract}
\newpage

\section{Introduction}

The origin of neutrino mass and mixing remains one of the most important questions in particle physics, along with other unresolved questions such as the matter-antimatter asymmetry of the Universe, and the possible unification of the strong and electroweak interactions.

The excess of matter over antimatter in the Universe has been firmly established. Although the standard model (SM) qualitatively satisfies the Sakharov conditions~\cite{Sakharov:1967dj}, the CP violation in SM is too small to explain the baryon asymmetry. A dynamical generation of baryon asymmetry requires new physics beyond SM. The discovery of neutrino masses and mixing in neutrino oscillation experiments is a great progress in particle physics. The smallness of neutrino masses can be naturally explained in the type I seesaw mechanism in which a number of heavy Majorana neutrinos are added to the SM, and a direct cosmological implication is leptogenesis~\cite{Fukugita:1986hr}. The observed baryon asymmetry can be produced by the out-of equilibrium, CP and $B-L$ violating decays of the right-handed neutrinos, see Refs.~\cite{Bodeker:2020ghk,DiBari:2021fhs} for recent review papers. In the type I seesaw model with three right-handed neutrinos, 18 additional parameters are introduced and the leptogenesis can successfully generate the observed baryon asymmetry if the neutrino Yukawa couplings and right-handed neutrino masses are freely chosen. It is difficult to test leptogenesis due to the large number of free parameters of in a generic seesaw model.

Masses and mixing angles of elementary fermions are known with good precision at present.  The mass eigenvalues of quarks and leptons have a large hierarchy, the flavor mixings in quark and lepton sectors are drastically different from each other: small quark mixing angles versus large lepton mixing angles. Understanding the origin of the flavor structure of SM is another of the greatest challenges in particle physics. The fundamental principal behind the observed patterns of fermion masses and mixing is still elusive,  and the approach of flavour symmetry has been extensively studied in the last several decades~\cite{King:2013eh,Feruglio:2017spp}. Many candidate groups for flavor symmetry have been discussed, and many models which can describe some pieces of the flavor puzzle were constructed. The flavor symmetry relates three generations of matter fields so that the resulting quark and lepton mass matrices have certain structure and the total number of free parameters in fermion mass matrices are reduced.

The flavor symmetry can not be exact symmetry and it has to be broken explicitly or spontaneously to accommodate the non-trivial mixing patterns of quarks and leptons. As a consequence, the fermion mass matrices are usually expanded as power series of the symmetry breaking parameters. The breaking of flavour symmetries typically relies on several scalar multiplets called flavons whose vacuum expectation values should be oriented along certain directions in flavor space. In addition, extra cyclic group generally is necessary to forbid the unwanted operators and realize the correct vacuum alignment~\cite{Feruglio:2017spp}. Thus the construction of flavon potential is intricate and make the flavor symmetry models rather complicated.

The modular invariance as flavor symmetry  is recently proposed in~\cite{Feruglio:2017spp}. From the top-down perspective, modular invariance can naturally arise from the compactification of a higher dimensional theory on a torus or an orbifold. In minimal schemes based on modular invariance, flavons are not necessary and the complex modulus $\tau$ is the unique symmetry breaking parameter, consequently the vacuum alignment problem is considerably simplified. The action of modular symmetry on the matter fields is characterized by the modular weights and their transformations under the finite modular group $\Gamma_N\equiv\Gamma/\Gamma(N)$, where $\Gamma(N)$ is the principal congruence subgroup of level $N$ of the modular group $\Gamma\equiv PSL(2, Z)$~\cite{Feruglio:2017spp}. Modular invariance constrains the Yukawa couplings to be level $N$ modular forms which are specific holomorphic functions of $\tau$. To be more general, one can use any normal subgroup of $\Gamma$ rather than $\Gamma(N)$ such that the Yukawa couplings are vector-valued modular forms of $SL(2, Z)$~\cite{Liu:2021gwa}. The modular invariance approach to the flavor puzzle, especially for the lepton flavor structure have been extensively exploited, and various modular invariant models have been constructed in past years with the groups $\Gamma_2\cong S_3$~\cite{Kobayashi:2018vbk,Kobayashi:2018wkl,Kobayashi:2019rzp,Okada:2019xqk}, $\Gamma_3\cong A_4$~\cite{Feruglio:2017spp,Criado:2018thu,Kobayashi:2018vbk,Kobayashi:2018scp,deAnda:2018ecu,Okada:2018yrn,Kobayashi:2018wkl,Novichkov:2018yse,Nomura:2019jxj,Okada:2019uoy,Nomura:2019yft,Ding:2019zxk,Okada:2019mjf,Nomura:2019lnr,Kobayashi:2019xvz,Asaka:2019vev,Gui-JunDing:2019wap,Zhang:2019ngf,Nomura:2019xsb,Wang:2019xbo,Kobayashi:2019gtp,King:2020qaj,Ding:2020yen,Okada:2020rjb,Nomura:2020opk,Asaka:2020tmo,Okada:2020brs,Yao:2020qyy,Feruglio:2021dte,Okada:2021qdf}, $\Gamma_4\cong S_4$~\cite{Penedo:2018nmg,Novichkov:2018ovf,deMedeirosVarzielas:2019cyj,Kobayashi:2019mna,King:2019vhv,Criado:2019tzk,Wang:2019ovr,Gui-JunDing:2019wap,Wang:2020dbp,Qu:2021jdy}, $\Gamma_5\cong A_5$~\cite{Novichkov:2018nkm,Ding:2019xna,Criado:2019tzk} and $\Gamma_7\cong PSL(2,Z_7)$~\cite{Ding:2020msi}.
Similar attempts have been made to construct modular invariant models for quarks and quark-lepton unification models based on the modular groups $A_4$~\cite{Okada:2018yrn,Okada:2019uoy,Okada:2020rjb,Yao:2020qyy}, $S_4$~\cite{Qu:2021jdy}, $T'$~\cite{Lu:2019vgm}, $S'_4$~\cite{Liu:2020akv} and $A'_5$~\cite{Yao:2020zml}. The formalism of modular invariance has been extended to include odd weight modular forms~\cite{Liu:2019khw}, fractional weight modular forms~\cite{Liu:2020msy,Yao:2020zml} and it can involve several moduli~\cite{deMedeirosVarzielas:2019cyj,Ding:2020zxw}. The interplay of modular symmetry and generalized CP symmetry has been studied~\cite{Novichkov:2019sqv,Baur:2019kwi,Baur:2019iai,Ding:2021iqp}. Moreover, modular invariance has been intensively discussed from a top-down perspective~\cite{Baur:2019kwi,Baur:2019iai,Nilles:2020nnc,Ishiguro:2020nuf,Baur:2020yjl,Almumin:2021fbk}.

The origin of all quark and lepton masses and mixing may be related with each other. This suggests to combine the Grand unified theories (GUTs) with flavor symmetry. The electromagnetic, weak, and strong forces are merged into a single force at high energies in GUTs. The quark and lepton fields are embedded into one or few gauge multiplets, thus
GUTs specifically predict relations among the fermion masses. Inclusion of flavor symmetry allows to predict the mixing parameters of quark and leptons. However, combining traditional flavor symmetry with GUTs also requires  the introduction of flavons and vacuum alignment in order to break the flavor symmetry, see Ref.~\cite{King:2017guk} for a review. Therefore there is a strong motivation for introducing modular symmetry in the context of GUTs such that the complication of flavor symmetry breaking can be removed and the predictive power of the GUTs models can be improved considerably. The possible combinations of $SU(5)$ GUTs and modular symmetry groups have been discussed in the literature, and several modular $SU(5)$ GUT models have been built at level 2~\cite{Kobayashi:2019rzp,Du:2020ylx}, level 3~\cite{deAnda:2018ecu,Chen:2021zty,Charalampous:2021gmf} and level 4~\cite{Zhao:2021jxg,King:2021fhl,Ding:2021zbg}.

Besides the simple $SU(5)$ GUT, the $SO(10)$ GUT is another prototype of grand unified theory. The unification of matter is even more complete in $SO(10)$ GUT, since the 16-dimensional spinor representation of $SO(10)$ can  accommodate all the known SM chiral fermions as well as the right handed neutrino. This makes $SO(10)$ GUTs particularly attractive candidates for explaining the origin of neutrino mass and mixing, the unification of the strong and electroweak interactions, and the matter-antimatter asymmetry of the universe, simultaneously. It is notable that both the right-handed neutrino mass matrix and neutrino Yukawa coupling are constrained by the $SO(10)$ GUT symmetry, and they are related with the quark and lepton mass matrices. The $SO(10)$ GUTs employing an conventional flavor symmetry has been studied so far. For example, models with $SO(10)\times A_4$~\cite{Morisi:2007ft,Grimus:2008tm,Bazzocchi:2008sp,Albaid:2009uv} and $SO(10)\times S_4$~\cite{Lee:1994qx,Hagedorn:2006ug,Dutta:2009bj,BhupalDev:2012nm,Bjorkeroth:2017ybg} have already been investigated. As in $SU(5)$ GUTs, the cumbersome vacuum alignment required in conventional flavor symmetry can be eliminated by the use of modular symmetry.

Following this line of reasoning, SUSY $SO(10)\times A_4$ modular models were constructed in~\cite{Ding:2021eva}. We found that the models involving in addition to the Higgs fields in the ${\mathbf{10}}$ and the ${\mathbf{\overline{126}}}$, also a Higgs field in the ${\mathbf{120}}$, proved to give a successful description of quark and lepton (including neutrino) masses and mixing, with many such models being found with sums of modular weights of up to 8 or less. The neutrino masses were generated by the type-I and/or type II seesaw mechanisms~\cite{Minkowski:1977sc,Yanagida:1979as,GellMann:1980vs,Mohapatra:1979ia,Schechter:1980gr} arising from the $SU(2)_L$ singlet and/or triplet components of the Higgs in the ${\mathbf{\overline{126}}}$ representation.
However the question of leptogenesis was not addressed for these models.

In this paper, we study the prediction for leptogenesis in two of the renormalizable SUSY $SO(10)\times A_4$ modular models from \cite{Ding:2021eva} in which the neutrino mass is dominantly generated by the type I seesaw mechanism
\footnote{The study of such models involving a mixture of type-I and type-II seesaw models is much more complicated and left for a future project.}. The evolution of the lepton asymmetries are described in terms of the three-flavored density matrix equations for three heavy Majorana neutrinos, where both vanishing initial condition and thermal initial condition of the right-handed neutrinos are considered. We also present an analytical approximation based on the Boltzmann equations. We find regions of parameter space compatible with the measured fermion masses and mixing parameters as well as the baryon asymmetry of the Universe. The predictions for the light neutrino masses, the effective mass in neutrinoless doble beta decay and the leptonic CP violation phases are discussed.

This paper is organized as follows. In section~\ref{sec:rev-modular-SO10} we briefly review the modular invariant $SO(10)$ models and the predictions for the fermion mass matrices. In section~\ref{section3} we present the formalism of density matrix to describe the evolution of the lepton asymmetries and the number densities of right-handed neutrinos. Moreover, the decay and washout of three right-handed neutrinos are considered in turns in the framework of Boltzmann equations, and an approximation formula for the final lepton asymmetry is given. In section~\ref{section4} we present our numerical analysis for the concerned benchmark models, the results of the fit
and the predictions of the models are given. Section~\ref{sec:conclusion} concludes the paper. In the appendix~\ref{sec:app-bad-examples} we give two examples where the analytical approximations disagree with the numerical results.

\section{\label{sec:rev-modular-SO10}Fermion masses in $SO(10)$ GUTs with $A_4$ modular symmetry}

It is known that all the fermionic multiplets of each SM generation, plus one right-handed neutrino singlet, fit exactly into the
16 dimensional spinorial representation of the $SO(10)$ grand unification group. As a consequence, in renormalizable $SO(10)$ models, the Higgs fields that contribute to fermion masses are in the $SO(10)$ representations
$\mathbf{10}$, $\overline{\mathbf{126}}$ and $\mathbf{120}$ which are denoted by $H$, $\overline{\Delta}$ and $\Sigma$ respectively. The most general form of the Yukawa superpotential for renormalizable $SO(10)$ is given by
\begin{equation}
\label{eq:Yukawa-SO10}\mathcal{W}_Y=\mathcal{Y}_{ij}^{10} \psi_{i} \psi_{j}H+\mathcal{Y}_{ij}^{\overline{126}}  \psi_{i} \psi_{j}\overline{\Delta}+ \mathcal{Y}_{ij}^{120}\psi_{i} \psi_{j} \Sigma\,,
\end{equation}
with $i=1, 2, 3$, and $\psi_i$ stand for the three generations of matter fields which are in the 16 dimensional representation of $SO(10)$. The Yukawa coupling matrices $\mathcal{Y}^{10}$ and $\mathcal{Y}^{\overline{126}}$ are symmetric in generation space, whereas $\mathcal{Y}_{ij}^{120}$ are  antisymmetric. Their explicit forms can be constrained by the flavor symmetry which relates the three generations of fermions. The GUT Higgs $H$ and $\overline{\Delta}$ contain one pair of up-type and down-type Higgs doublets while $\Sigma$ has two pairs of such doublets. It is assumed that only one linear combination of up-type and one of down-type Higgs doublets are light, and the remaining combinations acquire GUT scale masses. The vacuum expectation value (VEV) of the light Higgs doublets breaks the electroweak symmetry and generate the fermion masses. The mass matrices of quarks and  charged leptons can be written as~\cite{Dutta:2004zh,Dutta:2005ni,Grimus:2006bb,Altarelli:2010at,Joshipura:2011nn,Ding:2021eva}
\begin{eqnarray}
M_u &=&\left(\mathcal{Y}^{10} + r_2 \mathcal{Y}^{\overline{126}} +r_3 \mathcal{Y}^{120}\right)v_u,\nonumber\\
M_d &=& r_1 \left(\mathcal{Y}^{10}+ \mathcal{Y}^{\overline{126}} + \mathcal{Y}^{120}\right)v_d\,, \nonumber\\
\label{eq:mass-matrix-charged-fermions}M_{l} &=& r_1\left(\mathcal{Y}^{10}-3 \mathcal{Y}^{\overline{126}} + c_e \mathcal{Y}^{120}\right)v_d\,,
\end{eqnarray}
where $v_u$ and $v_d$ refer to the VEVs of the MSSM Higgs fields $H_u$ and $H_d$ respectively, and some ratios of VEVs have been absorbed into the Yukawa matrices $\mathcal{Y}^{10}$, $\mathcal{Y}^{\overline{126}}$ and $\mathcal{Y}^{120}$~\cite{Ding:2021eva}. Moreover, the Higgs multiplet $\overline{\Delta}$ also contains the Pati-Salam multiplets $(\overline{\mathbf{10}}, \mathbf{3}, \mathbf{1})$ and $(\mathbf{10}, \mathbf{1}, \mathbf{3})$ which can generate the Majorana masses for the left-handed and right-handed neutrinos. Hence generally the light neutrino mass receives contributions from both type I and type II seesaw mechanism,
\begin{equation}
M_{\nu}=M_{L}-M_DM^{-1}_RM^T_D\,,\label{eq:nu-matrix}
\end{equation}
with
\begin{equation}
\label{eq:nu-Dirac}M_{D} =\left(\mathcal{Y}^{10}-3 r_2 \mathcal{Y}^{\overline{126}} + c_\nu \mathcal{Y}^{120}\right)v_u,~~~M_R=v_R\mathcal{Y}^{\overline{126}},~~~M_L=v_L\mathcal{Y}^{\overline{126}}\,.
\end{equation}
The dimensionless parameters $r_{1,2,3}$ and $c_{e,\nu}$ in Eqs.~(\ref{eq:mass-matrix-charged-fermions},\ref{eq:nu-Dirac}) are determined by the Clebsch–Gordan coefficients of $SO(10)$ and the mixing among
the Higgs fields. In the present work, we consider the scenario of type I seesaw dominance which is the limit of $v_{L}\rightarrow0$.

\subsection{$SO(10)$ models with $A_4$ modular symmetry}

The $A_4$ group has twelve elements and it can be generated by two generators $S$ and $T$ obeying the relations
\begin{equation}
S^2=(ST)^3=T^3=1\,.
\end{equation}
It has four non-equivalent irreducible representations: one triplet $\mathbf{3}$ and three singlets $\mathbf{1}$, $\mathbf{1'}$ and $\mathbf{1''}$. The tensor product of two triplets is
$\mathbf{3}\otimes\mathbf{3}=\mathbf{1}\oplus\mathbf{1'}\oplus\mathbf{1''}\oplus\mathbf{3}_S\oplus\mathbf{3}_A$, where $\mathbf{3}_S$ and $\mathbf{3}_A$ denote the symmetric and the antisymmetric triplet combinations respectively. In terms of the components of the two triplets $\alpha=\left(\alpha_1, \alpha_2, \alpha_3\right)^T$ and $\beta=\left(\beta_1, \beta_2, \beta_3\right)^T$, we has
\begin{eqnarray}
\nonumber&&\alpha_1\beta_1+\alpha_2\beta_3+\alpha_3\beta_2\sim\mathbf{1}\,,\\
\nonumber&&\alpha_3\beta_3+\alpha_1\beta_2+\alpha_2\beta_1\sim\mathbf{1'}\,,\\
\nonumber&&\alpha_2\beta_2+\alpha_1\beta_3+\alpha_3\beta_1\sim\mathbf{1''}\,,\\
\nonumber&&\begin{pmatrix}
2\alpha_1\beta_1-\alpha_2\beta_3-\alpha_3\beta_2 \\
2\alpha_3\beta_3-\alpha_1\beta_2-\alpha_2\beta_1\\
2\alpha_2\beta_2-\alpha_1\beta_3-\alpha_3\beta_1
\end{pmatrix}\sim\mathbf{3}_S\,,\\
\label{eq:tensor-CG-A4}&&\begin{pmatrix}
\alpha_2\beta_3-\alpha_3\beta_2\\
\alpha_1\beta_2-\alpha_2\beta_1\\
\alpha_3\beta_1-\alpha_1\beta_3
\end{pmatrix}\sim\mathbf{3}_A\,.
\end{eqnarray}
There are three linearly independent modular forms of weight 2 at level 3~\cite{Feruglio:2017spp}, they transform in the triplet representation of $A_4$ up to the automorphy factor $(c\tau+d)^2$, and they can be explicitly constructed in terms of the Dedekind eta-function $\eta(\tau)$ as follow~\cite{Liu:2019khw}:
\begin{equation}
Y^{(2)}_{\mathbf{3}}(\tau)=
\begin{pmatrix}
\varepsilon^2(\tau) \\
\sqrt{2}\,\vartheta(\tau)\varepsilon(\tau) \\
-\vartheta^2(\tau)
\end{pmatrix}\equiv\begin{pmatrix}
Y_1(\tau) \\
Y_2(\tau) \\
Y_3(\tau)
\end{pmatrix}\,,
\end{equation}
with
\begin{equation}
\vartheta(\tau)=3\sqrt{2}\,\frac{\eta^3(3\tau)}{\eta(\tau)},\qquad \varepsilon(\tau)=-\frac{3\eta^3(3\tau)+\eta^3(\tau/3)}{\eta(\tau)}\,,
\end{equation}
and
\begin{equation}
\eta(\tau)=q^{1/24}\prod_{n=1}^\infty \left(1-q^n \right),\qquad q=e^{i2\pi\tau}\,.
\end{equation}
Then we can read out the $q$-expansion of the modular forms $Y_{1,2,3}(\tau)$ as
\begin{eqnarray}
\nonumber&&Y_1(\tau)=1 + 12q + 36q^2 + 12q^3 + 84q^4 + 72q^5+36q^6+96q^7+180q^8+12q^9+216q^{10}+\dots \,, \\
\nonumber&&Y_2(\tau)=-6q^{1/3}\left(1 + 7q + 8q^2 + 18q^3 + 14q^4+31q^5+20q^6+36q^7+31q^8+56q^9+32q^{10} +\dots\right)\,, \\
&&Y_3(\tau)=-18q^{2/3}\left(1 + 2q + 5q^2 + 4q^3 + 8q^4 +6q^5+14q^6+8q^7+14q^8+10q^9+21q^{10}+\dots\right)\,.
\end{eqnarray}
The higher weight modular forms of level 3 can be constructed from the tensor product of $Y^{(2)}_{\mathbf{3}}$. There are five linearly independent weight 4 modular forms which can be arrange into the $A_4$ multiplets $\mathbf{1}$, $\mathbf{1'}$ and $\mathbf{3}$, i.e.
\begin{eqnarray}
\nonumber Y^{(4)}_{\mathbf{1}}&=&(Y^{(2)}_{\mathbf{3}}Y^{(2)}_{\mathbf{3}})_{\mathbf{1}}=Y_1^2+2 Y_2 Y_3\,, \\
\nonumber Y^{(4)}_{\mathbf{1}'}&=&(Y^{(2)}_{\mathbf{3}}Y^{(2)}_{\mathbf{3}})_{\mathbf{1}'}=Y_3^2+2 Y_1 Y_2\,,\\
Y^{(4)}_{\mathbf{3}}&=&\frac{1}{2}(Y^{(2)}_{\mathbf{3}}Y^{(2)}_{\mathbf{3}})_{\mathbf{3}_S}=
\begin{pmatrix}
Y_1^2-Y_2 Y_3\\
Y_3^2-Y_1 Y_2\\
Y_2^2-Y_1 Y_3
\end{pmatrix}\,.
\end{eqnarray}
At weight 6, we have three modular multiplets which transform in $\mathbf{1}$ and $\mathbf{3}$ of $A_4$,
\begin{eqnarray}
Y^{(6)}_{\mathbf{1}}&=&(Y^{(2)}_{\mathbf{3}}Y^{(4)}_{\mathbf{3}})_{\mathbf{1}}=Y_1^3+Y_2^3+Y_3^3-3 Y_1 Y_2 Y_3\,,\nonumber\\
Y^{(6)}_{\mathbf{3}I}&=&Y^{(2)}_{\mathbf{3}}Y^{(4)}_{\mathbf{1}}=(Y_1^2+2Y_2Y_3)\begin{pmatrix}
Y_1\\
Y_2\\
Y_3
\end{pmatrix}\,,\nonumber\\
\label{eq:MF-w6}Y^{(6)}_{\mathbf{3}II}&=&Y^{(2)}_{\mathbf{3}}Y^{(4)}_{\mathbf{1}'}=
(Y_3^2+2 Y_1Y_2)\begin{pmatrix}
Y_3\\
Y_1\\
Y_2
\end{pmatrix}\,.
\end{eqnarray}

\begin{table}[t!]
\centering
\begin{tabular}{|c|c|c|c|c|} \hline  \hline
        &       $\psi_{1,2,3}$ & $H$ & $\Sigma$ & $\overline{\Delta}$ \\ \hline
$SO(10)$ & $\mathbf{16}$ & $\mathbf{10}$  &  $\mathbf{120}$  &  $\overline{\mathbf{126}}$ \\ \hline
$A_4$   & $\mathbf{3}$  &  $\mathbf{1}$  &  $\mathbf{1}$   &  $\mathbf{1}$ \\ \hline
$k_I$  &  $k_F$  &  $k_{10}$  &  $k_{120}$  &  $k_{\overline{126}}$     \\ \hline \hline
\end{tabular}
\caption{\label{tab:SO10-A4-modular-model}The transformation properties and modular weights of the $SO(10)$ matter fields and Higgs multiplets. }
\end{table}

Although the $SO(10)$ GUT symmetry unifies all the fermions of each generation into a single representation $\mathbf{16}$, the fermions in different generations are not related so that the Yukawa coupling matrices $\mathcal{Y}^{10}$, $\mathcal{Y}^{\overline{126}}$ and $\mathcal{Y}^{120}$ in Eq.~\eqref{eq:Yukawa-SO10} can be arbitrary complex symmetric and antisymmetric matrices respectively. In order to understand the observed flavor structure of quarks and leptons, the $A_4$ modular symmetry is imposed in the $SO(10)$ GUT. We assign the three generations of matter fields $\psi_{1,2,3}$ to a $A_4$ triplet since the flavor symmetry would effectively be the $Z_3$ subgroup generated by $T$ rather than $A_4$ for the singlet assignment. The GUT Higgs multiplets $H$, $\overline{\Delta}$ and $\Sigma$ all transform as $\mathbf{1}$. Taking into account the $A_4$ modular symmetry, the $SO(10)$ Yukawa coupling of Eq.~\eqref{eq:Yukawa-SO10} becomes
\begin{eqnarray}
\nonumber\mathcal{W}_{Y}&=&\sum_{\mathbf{r}_a}
\alpha_a\left(\psi\psi Y_{\mathbf{r}_a}^{(2k_F+k_{10})}(\tau)\right)_{\mathbf{1}}H+\sum_{\mathbf{r}_c}
\beta_b\left(\psi\psi Y_{\mathbf{r}_b}^{(2k_F+k_{120})}(\tau)\right)_{\mathbf{1}}\Sigma \\ \label{eq:W_Y}
&&~~~+\sum_{\mathbf{r}_c}\gamma_c\left(\psi\psi Y_{\mathbf{r}_c}^{(2k_F+k_{\overline{126}})}(\tau)\right)_{\mathbf{1}}\overline{\Delta}\,,
\end{eqnarray}
where $k_F$, $k_{10}$, $k_{\overline{126}}$ and $k_{120}$ are the modular weights of $\psi$, $H$, $\overline{\Delta}$ and $\Sigma$ respectively. Notice that one has to sum over the contributions of all independent modular multiplets at the relevant weights. Using the multiplication rules in Eq.~\eqref{eq:tensor-CG-A4}, the Yukawa matrices $\mathcal{Y}^{10}$ can be read off as follows,
\begin{eqnarray}
\nonumber&&\mathcal{Y}^{10}\Big|_{k=2k_F+k_{10}}=\alpha_1Y_{\mathbf1}^{(k)}(\tau)\left(
\begin{matrix}
 1 ~& 0 ~& 0 \\
 0 ~& 0 ~& 1 \\
 0 ~& 1 ~& 0 \\
\end{matrix}
\right)+\alpha_2Y_{\mathbf1'}^{(k)}(\tau)\left(
\begin{matrix}
 0 ~& 0 ~& 1 \\
 0 ~& 1 ~& 0 \\
 1 ~& 0 ~& 0 \\
\end{matrix}
\right)\\
&&\qquad +\alpha_3Y_{\mathbf1''}^{(k)}(\tau)\left(
\begin{matrix}
 0 ~& 1 ~& 0 \\
 1 ~& 0 ~& 0 \\
 0 ~& 0 ~& 1 \\
\end{matrix}
\right)+\alpha_4\left(\begin{matrix}
 2Y^{(k)}_{{{\mathbf3}},1}(\tau) ~&~ -Y^{(k)}_{{{\mathbf3}},3}(\tau)  ~&~ -Y^{(k)}_{{{\mathbf3}},2}(\tau) \\
 -Y^{(k)}_{{{\mathbf3}},3}(\tau) ~&~ 2Y^{(k)}_{{{\mathbf3}},2}(\tau)  ~&~ -Y^{(k)}_{{{\mathbf3}},1}(\tau) \\
 -Y^{(k)}_{{{\mathbf3}},2}(\tau) ~&~ -Y^{(k)}_{{{\mathbf3}},1}(\tau)  ~&~ 2Y^{(k)}_{{{\mathbf3}},3}(\tau)
\end{matrix}
\right)\,,
\end{eqnarray}
and $\mathcal{Y}^{\overline{126}}$ takes a similar form with $\alpha_i$ replaced by $\gamma_i$. Because the coupling with $\Sigma$ is antisymmetric in the generation indices, only the antisymmetric triplet contraction $(\psi\psi)_{\mathbf{3}_A}$ contributes to the Yukawa matrix $\mathcal{Y}^{120}$,
\begin{eqnarray}
\mathcal{Y}^{120}\Big|_{k=2k_F+k_{120}}=\beta_1\left(\begin{matrix}
 0 ~&~ Y^{(k)}_{{{\mathbf3}},3}(\tau)  ~&~ -Y^{(k)}_{{{\mathbf3}},2}(\tau) \\
 -Y^{(k)}_{{{\mathbf3}},3}(\tau) ~&~ 0  ~&~ Y^{(k)}_{{{\mathbf3}},1}(\tau) \\
 Y^{(k)}_{{{\mathbf3}},2}(\tau) ~&~ -Y^{(k)}_{{{\mathbf3}},1}(\tau)  ~&~ 0
\end{matrix}\right)\,.
\end{eqnarray}

\subsection{\label{subsec:benchmark-models} Benchmark models }

In the following, we present two typical $SO(10)\times A_4$ modular models given in our previous work~\cite{Ding:2021eva}, then study the predictions for leptogenesis. The first benchmark model denoted as \texttt{BM1} is characterized by the modular weights $(2k_F+k_{10}\,, 2k_F+k_{120}\,, 2k_F+k_{\overline{126}})=(4,2,4)$, and it is the non-minimal model 2 of~\cite{Ding:2021eva}. Thus the modular invariant superpotential is given by
\begin{eqnarray}
\nonumber\mathcal{W}_Y &=&
\alpha_1 Y_{\mathbf{1}}^{(4)} \psi \psi H
+\alpha_2 Y_{\mathbf{1}'}^{(4)} \psi \psi H
+\alpha_3 Y_{\mathbf{3}}^{(4)} \psi \psi H
+\beta_1 Y_{\mathbf{3}}^{(2)} \psi \psi \Sigma \\ \label{eq:SP_non_minimal_model_2}
&&+\gamma_1 Y_{\mathbf{1}}^{(4)} \psi \psi \overline{\Delta}
+\gamma_2 Y_{\mathbf{1}'}^{(4)} \psi \psi \overline{\Delta}
+\gamma_3 Y_{\mathbf{3}}^{(4)} \psi \psi \overline{\Delta}\,.
\end{eqnarray}
The second model denoted as \texttt{BM2} is specified by the modular weights $(2k_F+k_{10}\,, 2k_F+k_{120}\,, 2k_F+k_{\overline{126}})=(4,6,4)$, and it is the non-minimal model 3 of~\cite{Ding:2021eva}. We can straightforwardly read out the superpotential as follows,
\begin{eqnarray}
\nonumber\mathcal{W}_Y &=&
\alpha_1 Y_{\mathbf{1}}^{(4)} \psi \psi H
+\alpha_2 Y_{\mathbf{1}'}^{(4)} \psi \psi H
+\alpha_3 Y_{\mathbf{3}}^{(4)} \psi \psi H+\beta_1 Y_{\mathbf{3}I}^{(6)} \psi \psi\Sigma \\ \label{eq:SP_non_minimal_model_3}
&&+\beta_2 Y_{\mathbf{3}II}^{(6)} \psi \psi\Sigma +\gamma_1 Y_{\mathbf{1}}^{(4)} \psi \psi \overline{\Delta}
+\gamma_2 Y_{\mathbf{1}'}^{(4)} \psi \psi \overline{\Delta}
+\gamma_3 Y_{\mathbf{3}}^{(4)} \psi \psi \overline{\Delta} \,.
\end{eqnarray}
In the case that the light neutrino mass is dominated by the type-I seesaw contribution, the best fit values of the free parameters for the above two models are listed in table~\ref{tab:fit-non_minimal_model_2-VIA} and table~\ref{tab:fit-non_minimal_model_2_second-TIA}, and we see that the experimental data can be accommodated very well.

\section{Prediction for baryon asymmetry via leptogenesis}
\label{section3}

The baryon asymmetry of the Universe is a puzzle in particle physics, and its  value can be inferred from observations in two independent ways. The first is from the big bang nucleosynthesis. With the assumption of only three light
neutrinos, the predictions for the abundances of the light elements  D, $^3{\rm He}$, $^4{\rm He}$, and
$^7${\rm Li} depend on a single parameter, that is the
difference between the number of baryons and anti-baryons normalized
to the number of photons:
\begin{equation}
\eta\equiv \frac{n_B- n_{\bar{B}}}{n_{\gamma }} {\Big |}_{0},
\label{eq-08:etaB}
\end{equation}
where the subscript $0$ means ``at present time''. The range of $\eta$ which is consistent with the abundances of D and $^4{\rm He}$ is determined to be~\cite{ParticleDataGroup:2020ssz}
\begin{equation}
\eta =(6.143 \pm 0.190) \times 10^{-10}\,,
\end{equation}
at $1\sigma$ confidence level. This range corresponds to
\begin{equation}
\label{eq:YB-BBN} Y_{\Delta B}=(8.72\pm 0.27)\times 10^{-11}\,.
\end{equation}
Notice that $Y_{\Delta B}$ is defined as $Y_{\Delta B} \equiv  \frac{n_B- n_{\bar{B}}}{s} {\Big |}_{0}$ and it is related to the parameter $\eta$ through $Y_{\Delta B}\simeq\eta/7.04$. Furthermore the baryon asymmetry produces acoustic oscillations in the power spectrum of the cosmic microwave background~\cite{Planck:2018vyg}. Observing these oscillations gives an even tighter bound on the value of baryon asymmetry,
\begin{equation}
Y_{\Delta B}=\left(8.703 \pm 0.113\right)\times~10^{-11} ~~~(95\%, \text{TT,TE,EE+lowE})
\end{equation}
It is impressive that the above two unrelated measurement approaches give consistent range of baryon asymmetry. In this work, we will use the conservative value of Eq.~\eqref{eq:YB-BBN} from the big bang nucleosynthesis.

Generally speaking, both type I and type II seesaw mechanisms contribute to the light neutrino mass in $SO(10)$ GUT, the contribution of type I seesaw is assumed to be dominant over that of type II in the present work. Then the CP violating out-of-equilibrium decay of the heavy Majorana neutrinos can generate a lepton asymmetry which is subsequently partially converted to a baryon asymmetry through the $B+L$ violating sphaleron processes~\cite{Fukugita:1986hr}. In this section, we shall study whether the measured value of the baryon asymmetry of the Universe $Y_{\Delta B}=(8.72\pm 0.27)\times 10^{-11}$ in Eq.~\eqref{eq:YB-BBN} can be correctly generated through thermal leptogenesis in the above benchmark $SO(10)\times A_4$ modular models. The evolution of the lepton asymmetry is described by the density matrix equation.

\subsection{Density matrix equation }

In the simplest scenario of leptogenesis, the lepton and anti-lepton final states from the decays of the heavy neutrino $N_i$ are in coherent superposition of flavor,
\begin{equation}
\label{eq:lepton-state-l2-l3}|\ell_i\rangle=\sum_\alpha C_{i\alpha}|\ell_\alpha\rangle\,,~~~~|\bar{\ell}_i\rangle=\sum_\alpha \bar{C}_{i\alpha}|\bar{\ell}_\alpha\rangle\,,
\end{equation}
where $i=1, 2, 3$ and $\alpha=e, \mu, \tau$, and the projection coefficients at tree-level are given by
\begin{equation}
\label{eq:C-par}C_{i\alpha}=\bar{C}_{i\alpha}=\dfrac{(\lambda_\nu)_{\alpha i}}{\sqrt{(\lambda_\nu^\dag\lambda_\nu)_{ii}}}\,.
\end{equation}
Notice that we work in the basis where both charged lepton and right-handed neutrino mass matrices are diagonal. This is a good approximation for high temperature $T\gg 10^{12}$ GeV when the charged lepton Yukawa interactions are negligible. However, the coherent evolution of the states $|\ell_i\rangle$ and $|\bar{\ell}_i\rangle$ would be broken down and different lepton flavors are distinguishable when these interactions are in thermal equilibrium. Then the left-handed leptons can be rapidly converted to right-handed components through scattering with the Higgs doublet. To be more specific, the $\tau$ and $\mu$ Yukawa interactions come into thermal equilibrium at the temperature $T\sim 10^{12}$ GeV and $T\sim 10^{9}$ GeV respectively.

It has been shown that the density matrix equation can provide an accurate description for the time evolution of the lepton asymmetry~\cite{Nardi:2006fx,Abada:2006ea,Barbieri:1999ma,Abada:2006fw,DeSimone:2006nrs,Blanchet:2006ch,Blanchet:2011xq}.  The approach of the density matrix accounts for the quantum decoherence effect and it allows to quantitatively describe the transitions between the one-flavored, two-flavored and three-flavored regimes. Comprehensive analysis of leptogenesis has been performed in the framework of density matrix~\cite{Moffat:2018wke,Moffat:2018smo,Granelli:2021fyc}. The formalism of the density matrix for leptogenesis has been given in Ref.~\cite{Blanchet:2011xq}, we shall follow closely the notation of~\cite{Blanchet:2011xq} in the following. For three decaying heavy Majorana neutrinos, the most general form of the density matrix equations are given by
\begin{eqnarray}
\nonumber\dfrac{\mathrm{d}N_{N_i}}{\mathrm{d}z}&=& -D_i(N_{N_i}-N^{\rm eq}_{N_i})\,, \\
\dfrac{\mathrm{d}N^{B-L}_{\alpha\beta}}{\mathrm{d}z}&=&\sum_i\left[\varepsilon^{(i)}_{\alpha\beta}\nonumber D_i(N_{N_i}-N^{\rm eq}_{N_i})-\dfrac{1}{2}W_i\left\{P^{0(i)},N^{B-L}\right\}_{\alpha\beta}\right]\\
\nonumber &-&\dfrac{\Im(\Lambda_\tau)}{Hz}\left(\delta_{\alpha\tau}N^{B-L}_{\tau\beta}+\delta_{\beta\tau}N^{B-L}_{\alpha\tau}-2\delta_{\alpha\tau}\delta_{\beta\tau}N^{B-L}_{\tau\tau}\right) \\
&-&\dfrac{\Im(\Lambda_\mu)}{Hz}\left(\delta_{\alpha\mu}N^{B-L}_{\mu\beta}+\delta_{\beta\mu}N^{B-L}_{\alpha\mu}-2\delta_{\alpha\mu}\delta_{\beta\mu}N^{B-L}_{\mu\mu}\right)\,,\label{eq:dme}
\end{eqnarray}
with $z=M_1/T$. The projection matrices are
\begin{equation}
P^{0(i)}_{\alpha\beta}=C_{i\alpha}C^{*}_{i\beta}\,,
\end{equation}
which describe how a given flavor of lepton is washed out. The quantity $N_{N_i}$ is the number of the heavy neutrino $N_i$ in a portion of comoving volume which contains one photon at the temperature $z\simeq0$. Hence the equilibrium number density $N^{\rm eq}_{N_i}$ for the Boltzmann statistics is given by
\begin{equation}
N^{\rm eq}_{N_i}(z)=\dfrac{3}{8}x_iz^2\mathcal{K}_2(z_i)\,,
\end{equation}
where $x_i=M_i^2/M_1^2$, $z_i=\sqrt{x_i}z$, and $\mathcal{K}_2(z)$ the modified Bessel functions of the second kind. Consequently we have $N^{\rm eq}_{N_i}(z\simeq 0)=3/4$. The diagonal entries $N^{B-L}_{\alpha\alpha}$ are the known number densities for the $B/3-L_{\alpha}$ asymmetry, and the off-diagonal elements $N^{B-L}_{\alpha\beta}$ describe the degree of coherence between the flavor states. The rescaled decay rate $D_i$ of the right-handed neutrino $N_i$ is given by
\begin{equation}
D_i(z)=K_ix_iz\dfrac{\mathcal{K}_1(z_i)}{\mathcal{K}_2(z_i)}\,,~~~K_i=\dfrac{M_i(\lambda_\nu^\dag\lambda_\nu)_{ii}}{8\pi H(T=M_i)}\,,
\end{equation}
where $H$ is the Hubble expansion rate. The washout parameters $W_i$ read
\begin{equation}
W_i(z)=\frac{1}{4}K_i \sqrt{x_i}\,z_i^3\,{\cal K}_1(z_i) \,.
\end{equation}
Furthermore, the last two terms in Eq.~\eqref{eq:dme} describe the effect of charged lepton Yukawa interactions which can induce the transition of left-handed leptons to right-handed leptons. They describe decoherence via the damping of the off-diagonal terms of the $B-L$ asymmetry matrix $N^{B-L}$ and they are determined as
\begin{eqnarray}
\frac{\Im(\Lambda_\mu)}{H\, z} & = & \frac{8 \times 10^{-3}\, y_\mu^2\, T}{H z}=1.72\times 10^{-10}\,\frac{M_\text{Pl}}{M_1}\,,\nonumber \\
\frac{\Im(\Lambda_\tau)}{H\, z} & = & \frac{8 \times 10^{-3}\, y_\tau^2\, T}{H z}=4.87\times 10^{-8}\,\frac{M_\text{Pl}}{M_1}\,,
\end{eqnarray}
where $M_\text{Pl}\simeq1.22\times10^{19}$ GeV is the Planck mass, the Yukawa couplings $y_{\mu}\simeq6.07\times 10^{-4}$ and $y_{\tau}\simeq1.02\times 10^{-2}$ are fixed by the $\mu$ and $\tau$ masses $m_{\mu}\simeq105.66$ MeV and $m_{\tau}\simeq1776.86$ MeV respectively. When the temperature goes below $10^{12}$ GeV, the $\tau$ Yukawa interactions come into thermal equilibrium and leads to decoherence of $\tau$ lepton states. This implies the transition from an unflavoured to two-flavoured regime. When the temperature drops below $10^9$ GeV the similar effects arise from the $\mu$ Yukawa interactions. Analogously, the electron Yukawa dependent damping term should be considered for $M_{1}<10^5$ GeV. Finally the CP asymmetry $\varepsilon_{\alpha \beta}^{(i)}$ generated by $N_i$ decay is of the following form~\cite{Covi:1996wh,DeSimone:2006nrs,Abada:2006ea,Beneke:2010dz,Blanchet:2011xq}
\begin{eqnarray}
\varepsilon_{\alpha \beta}^{(i)} & = &  \frac{3}{32 \pi \left(\lambda^\dagger_{\nu}\lambda_{\nu} \right)_{ii}}\,\sum_{j\neq i}\bigg\{i\left[(\lambda_{\nu})_{\alpha i} (\lambda^*_{\nu})_{\beta j} (\lambda^\dagger \lambda)_{ji} - (\lambda^*_{\nu})_{\beta i} (\lambda_{\nu})_{\alpha j} (\lambda^\dagger_{\nu} \lambda_{\nu})_{ij} \right] f_1\left(\frac{x_j}{x_i}\right)\nonumber  \\
&+&   i\left[ (\lambda_{\nu})_{\alpha i} (\lambda^*_{\nu})_{\beta j} (\lambda^\dagger_{\nu}\lambda_{\nu})_{ij} - (\lambda^*_{\nu})_{\beta i} (\lambda_{\nu})_{\alpha j} (\lambda^\dagger_{\nu} \lambda_{\nu})_{ji} \right]f_2\left(\frac{x_j}{x_i}\right) \bigg\}\,,
\end{eqnarray}
with
\begin{equation}
f_1(x)=\frac{2\sqrt{x}}{3}\left[(1+x)\,\ln\left( \frac{1+x}{x}\right) - \frac{2-x}{1-x}\right],~~~f_2(x)=\frac{2}{3(x-1)}\,.
\end{equation}
The diagonal components $\varepsilon^{(i)}_{\alpha\alpha}$ of the CP asymmetry matrix are exactly the usual flavored CP asymmetries,
\begin{equation}
\varepsilon_{\alpha \alpha}^{(i)} =  \frac{3}{16 \pi \left(\lambda^\dagger_{\nu}\lambda_{\nu} \right)_{ii}}\,\sum_{j\neq i}\bigg\{\Im\left[(\lambda^*_{\nu})_{\alpha i} (\lambda_{\nu})_{\alpha j} (\lambda^\dagger_{\nu} \lambda_{\nu})_{ij} \right] f_1\left(\frac{x_j}{x_i}\right)  + \Im\left[(\lambda^*_{\nu})_{\alpha i} (\lambda_{\nu})_{\alpha j} (\lambda^\dagger_{\nu} \lambda_{\nu})_{ji} \right]f_2\left(\frac{x_j}{x_i}\right) \bigg\}\,,
\end{equation}
while the off-diagonal components $\varepsilon^{(j)}_{\alpha\beta}=(\varepsilon^{(j)}_{\beta\alpha})^*$ and they are not necessarily real~\cite{Blanchet:2011xq}. The trace of the matrix
$N^{B-L}_{\alpha\beta}$ gives the total lepton asymmetry
\begin{equation}
N^{\rm f}_{B-L} = \sum_\alpha N^{B-L}_{\alpha \alpha}\,,
\end{equation}
which is then converted into the baryon asymmetry of the Universe,
\begin{equation}
\eta= \dfrac{28}{79}\dfrac{1}{27}\,N^{\rm f}_{B-L}\simeq0.013\, N^{\rm f}_{B-L}\,,
\end{equation}
where the factor $28/79$ accounts for the partial conversion of the
$B-L$ asymmetry into the baryon asymmetry by sphaleron process, and the second factor $1/27$ describes the dilution of the baryon asymmetry due to the change of the photon density between leptogenesis and recombination. In the following numerical analysis, we will use the Python package ULYSSES~\cite{Granelli:2020pim} to solve the coupled density matrix equations shown in Eq.~\eqref{eq:dme}.

\subsection{\label{subsec:analytical-app}Analytic approximation within the Boltzmann equations}

In our benchmark models,  the right-handed neutrino masses are hierarchical with $M_3\gg M_2\gtrsim10^{12}\,{\rm GeV}$ while $10^9\,{\rm GeV}\ll M_1\ll 10^{12}\,{\rm GeV}$. Therefore the decays of $N_3$ and $N_2$ are in the single-flavored regime while $N_1$ decay is in the two-flavored regime. In the following, we shall analyze the sequential decay of the three heavy Majorana neutrinos and evolution of lepton symmetry in terms of Boltzmann equations. In the first stage around the temperature $T\sim M_3$, a $B-L$ asymmetry is generated from the $N_3$ decay. In the second (third) stage at $T\sim M_2(M_1)$ another $B-L$ asymmetry is generated from the $N_2 (N_1)$ decays, while the inverse decay process of $N_2 (N_1)$ become effective and washout the asymmetry of $N_3$ decay to some level. In each stage we can use the Boltzmann equations to obtain an approximate analytical expression for the $B-L$ asymmetry.

In the first stage for $M_3\gtrsim T\gtrsim T_{B_3}$, where $T_{B_3}\simeq M_3/z_{B_3}$ is the freeze-out temperature of the $N_3$ inverse decay and $z_{B_3}\simeq 2+4K_3^{0.13}\,e^{-2.5/K_3}=\mathcal{O}(1-10)$~\cite{Blanchet:2006dq}, the rates of the $\tau$ and $\mu$ Yukawa interactions are much smaller than the expansion rate of the Universe. Consequently the charged lepton lepton state produced from $N_3$ decay is in coherent superposition and the flavor states are indistinguishable. In the single flavor approximation, the time evolution of the number densities of $N_3$ and $B-L$ can be described by the following semi-classical Boltzmann equations:
\begin{eqnarray}
\nonumber \dfrac{{\rm d}N_{N_3}}{{\rm d}z_3}&=&-D_3(N_{N_3}-N^{\rm eq}_{N_3})\,, \\
\dfrac{{\rm d}N_{\Delta_3}}{{\rm d}z_3}&=&\varepsilon_3 D_3(N_{N_3}-N^{\rm eq}_{N_3})-W_3N_{\Delta_3} \,.
\end{eqnarray}
The solution to the above equations is given by~\cite{Buchmuller:2004nz}
\begin{eqnarray}
N_{\Delta_3}({T\simeq T_{B_3}})=\dfrac{3}{4}\varepsilon_3\kappa(K_3)\,.
\end{eqnarray}
where $\kappa(K_3)$ is the so-called efficiency factor and its value depends on the initial condition of decaying heavy neutrinos. For thermal initial abundance of right-handed neutrinos, it is well approximated by~\cite{Buchmuller:2004nz},
\begin{eqnarray}
\label{eq:kappa-thermal}\kappa(x)=\dfrac{2}{x z_B(x)}\left[1-\exp\left(-\dfrac{1}{2}xz_B(x)\right)\right]\,.
\end{eqnarray}
For the vanishing initial abundance of right-handed neutrinos, the approximation formula for the efficiency factor is~\cite{Buchmuller:2004nz}
\begin{eqnarray}
\nonumber\kappa(K)&=&\kappa^-(K)+\kappa^+(K)\,, \\
\nonumber \kappa^-(K)&=&-2\,e^{-\frac{2}{3}N(K)}\left[e^{\frac{2}{3}\overline{N}(K)}-1\right]\,, \\
\label{eq:kappa-vanishing}\kappa^+(K)&=&\dfrac{2}{z_B(K)K}\left[1-e^{-\frac{2}{3}z_B(K)K\overline{N}(K)}\right]\,,
\end{eqnarray}
with
\begin{eqnarray}
N(K)=\dfrac{9\pi K}{16}\,, ~~~~~\overline{N}(K)=\dfrac{N(K)}{
\left(1+\sqrt{\frac{N(K)}{3/4}}\right)^2} \,.
\end{eqnarray}
The lepton state $|\ell_{3}\rangle$ produced by $N_3$ decay can be regarded as coherent superposition of a $\ell_2$ parallel component and of a $\ell_2$ orthogonal component which is denoted as $|\ell_{\tilde{2}}\rangle$. To be more concrete, $|\ell_{\tilde{2}}\rangle$ is the projection of $|\ell_3\rangle$ on the plane orthogonal to $|\ell_2\rangle$ such that $\langle\ell_{\tilde{2}}|\ell_2\rangle=0$, and its explicit expression is
\begin{eqnarray}
|\ell_{\tilde{2}}\rangle=\dfrac{1}{\sqrt{1-p_{32}}}(|\ell_3\rangle-\langle\ell_2|\ell_3\rangle|\ell_2\rangle)\,,
\end{eqnarray}
where the probabilities $p_{32}\equiv |\langle\ell_2|\ell_3\rangle|^2$ and $p_{3\tilde{2}}\equiv |\langle\ell_{\tilde{2}}|\ell_3\rangle|^2$ with $p_{32}+p_{3\tilde{2}}=1$. From Eq.~\eqref{eq:lepton-state-l2-l3}, we can read out the general expression of $p_{ij}$ as
\begin{eqnarray}
p_{ij}=\Big|\sum_\alpha C_{i\alpha}C_{j\alpha}^*\Big|^2&=&\dfrac{|(\lambda_\nu^\dag\lambda_\nu)_{ij}|^2}{(\lambda_\nu^\dag\lambda_\nu)_{ii}(\lambda_\nu^\dag\lambda_\nu)_{jj}}\,.
\end{eqnarray}
Correspondingly the same decomposition can be made for the asymmetry $N_{\Delta_3}$,
\begin{eqnarray}
\nonumber N_{\Delta_2}({T\simeq T_{B_3}})&=&p_{32}N_{\Delta_3}({T\simeq T_{B_3}})=\dfrac{3}{4}p_{32}\varepsilon_3\kappa(K_3) \,, \\
N_{\Delta_{\tilde{2}}}({T\simeq T_{B_3}})&=&p_{3\tilde{2}}N_{\Delta_3}({T\simeq T_{B_3}})=\dfrac{3}{4}(1-p_{32})\varepsilon_3\kappa(K_3) \,.
\end{eqnarray}
These two asymmetries can be used as initial condition at the beginning of the $N_2$ production and decay. When the temperature goes down to $T\sim M_2$, the $N_2$ decay and inverse processes break the coherent evolution of $|\ell_3\rangle$ which becomes an incoherent mixture of $|\ell_2\rangle$ and $|\ell_{\tilde{2}}\rangle$. The $|\ell_2\rangle$ component undergoes the washout from $N_2$ while the orthogonal component $|\ell_{\tilde{2}}\rangle$ is unwashed. The kinetic evolution equations are given by
\begin{eqnarray}
\nonumber \dfrac{{\rm d}N_{N_2}}{{\rm d}z_2}&=&-D_2(N_{N_2}-N^{\rm eq}_{N_2})\,, \\
\nonumber \dfrac{{\rm d}N_{\Delta_2}}{{\rm d}z_2}&=&\varepsilon_2 D_2(N_{N_2}-N^{\rm eq}_{N_2})-W_2N_{\Delta_2} \,, \\
\label{eq:N2-Boltzmann}\dfrac{{\rm d}N_{\Delta_{\tilde{2}}}}{{\rm d}z_2}&=&0 \,.
\end{eqnarray}
Hence the asymmetries at the temperature $T\simeq T_{B_2}$ take the form,
\begin{eqnarray}
\nonumber N_{\Delta_2}({T\simeq T_{B_2}})&=&\dfrac{3}{4}\varepsilon_2\kappa(K_2)+N_{\Delta_2}({T\simeq T_{B_3}})\,e^{-\frac{3\pi}{8}K_2} \,, \\
N_{\Delta_{\tilde{2}}}({T\simeq T_{B_2}})&=&N_{\Delta_{\tilde{2}}}({T\simeq T_{B_3}}) \,,
\end{eqnarray}
Since $10^9\,{\rm GeV}\ll M_1 \ll 10^{12}\,{\rm GeV}$, the $\tau$ Yukawa interaction becomes effective before the onset of $N_1$ washout processes.
Let us indicate with $M_1\ll T_\star\ll 10^{12}\,{\rm GeV}$ that value of temperature below which one can approximate the coherence of $|\tau\rangle$ and the orthogonal component completely damped. Thus the lepton quantum state
can be treated as an incoherent mixture of three components: $|\tau\rangle$, $|\tilde{\tau}_2\rangle$ and $|\tilde{\tau}_{\tilde{2}}\rangle$, where $|\tilde{\tau}_2\rangle$ and $|\tilde{\tau}_{\tilde{2}}\rangle$ are projections of $|\ell_2\rangle$ and $|\ell_{\tilde{2}}\rangle$ on the $e-\mu$ plane, respectively. Hence the lepton states $|\tilde{\tau}_2\rangle$ and $|\tilde{\tau}_{\tilde{2}}\rangle$ read as
\begin{eqnarray}
\nonumber|\tilde{\tau}_2\rangle&=&\dfrac{1}{\sqrt{1-p_{2\tau}}}\left(|\ell_2\rangle-\langle\tau|\ell_2\rangle|\tau\rangle\right)=\dfrac{1}{\sqrt{1-p_{2\tau}}}\left(C_{2\mu}|\mu\rangle+C_{2e}|e\rangle\right)\,, \\
|\tilde{\tau}_{\tilde{2}}\rangle&=&\dfrac{1}{\sqrt{1-p_{\tilde{2}\tau}}}\left(|\ell_{\tilde{2}}\rangle-\langle\tau|\ell_{\tilde{2}}\rangle|\tau\rangle\right)\,.
\end{eqnarray}
The corresponding probabilities are $p_{2\tau}=|\langle \tau|\ell_2\rangle|^2=|C_{2\tau}|^2$, $p_{2\tilde{\tau}_2}=|\langle \tilde{\tau}_2|\ell_2\rangle|^2$, $p_{\tilde{2}\tau}=|\langle\tau|\ell_{\tilde{2}}\rangle|^2$ and $p_{\tilde{2}\tilde{\tau}_{\tilde{2}}}=|\langle \tilde{\tau}_{\tilde{2}}|\ell_{\tilde{2}}\rangle|^2$ with
\begin{eqnarray}
\nonumber p_{\tilde{2}\tau}&=&|\langle\tau|\ell_{\tilde{2}}\rangle|^2=\dfrac{1}{1-p_{32}}\Big[p_{3\tau}+p_{32}p_{2\tau}-2\Re\Big(\sum_{\alpha} C_{3\tau}C^*_{2\tau}C_{2\alpha}C^*_{3\alpha}\Big)\Big]\,,\\
&&\hskip-0.35in p_{2\tau}+p_{2\tilde{\tau}_2}=p_{\tilde{2}\tau}+p_{\tilde{2}\tilde{\tau}_{\tilde{2}}}=1\,.
\end{eqnarray}
Thus the $B-L$ asymmetry before $N_1$ decay can be decomposed into the following three parts:
\begin{eqnarray}
\nonumber N_{\Delta_\tau}(T\simeq T_\star)&=&p_{2\tau}N_{\Delta_2}(T\simeq T_{B_2})+p_{\tilde{2}\tau}N_{\Delta_{\tilde{2}}}(T\simeq T_{B_2})\,, \\
\nonumber N_{\Delta_{\tilde{\tau}_2}}(T\simeq T_\star)&=&(1-p_{2\tau})N_{\Delta_2}(T\simeq T_{B_2})\,, \\
N_{\Delta_{\tilde{\tau}_{\tilde{2}}}}(T\simeq T_\star)&=&(1-p_{\tilde{2}\tau})N_{\Delta_{\tilde{2}}}(T\simeq T_{B_2})\,.
\end{eqnarray}
The asymmetries in the tau flavor and the orthogonal components  $|\tilde{\tau}_1\rangle$  and  $|\tilde{\tau}_{\tilde{1}}\rangle$ experience different washout from inverse decay, The quantum state $|\tilde{\tau}_1\rangle$ is the projection of $|\ell_1\rangle$ on the $e-\mu$ plane and $|\tilde{\tau}_{\tilde{1}}\rangle$ is the quantum state orthogonal to $|\tilde{\tau}_1\rangle$ in $e-\mu$ plane,
\begin{eqnarray}
\nonumber |\tilde{\tau}_1\rangle&=&\dfrac{1}{\sqrt{1-p_{1\tau}}}\left(|\ell_1\rangle-\langle\tau|\ell_1\rangle|\tau\rangle\right)=\dfrac{1}{\sqrt{1-p_{1\tau}}}\left(C_{1\mu}|\mu\rangle+C_{1e}|e\rangle\right) \,, \\
|\tilde{\tau}_{\tilde{1}}\rangle&=&\dfrac{1}{\sqrt{1-p_{1\tau}}}\left(-C^*_{1e}|\mu\rangle+C^*_{1\mu}|e\rangle\right)\,,
\end{eqnarray}
with $p_{1\tau}=|C_{1\tau}|^2$. One can use the two-flavored Boltzmann equations to describe the time evolution of the asymmetries during $N_1$ leptogenesis,
\begin{eqnarray}
\nonumber \dfrac{{\rm d}N_{N_1}}{{\rm d}z_1}&=&-D_1(N_{N_1}-N^{\rm eq}_{N_1})\,, \\
\nonumber \dfrac{{\rm d}N_{\Delta_\tau}}{{\rm d}z_1}&=&\varepsilon_{1\tau}D_1(N_{N_1}-N^{\rm eq}_{N_1})-p_{1\tau}W_1N_{\Delta_\tau}\,, \\
\nonumber \dfrac{{\rm d}N_{\Delta_{\tilde{\tau}_1}}}{{\rm d}z_1}&=&\varepsilon_{1\tilde{\tau}_1}D_1(N_{N_1}-N^{\rm eq}_{N_1})-p_{1\tilde{\tau}_1}W_1N_{\Delta_{\tilde{\tau}_1}}\,, \\
\label{eq:N1-Boltzmann}\dfrac{{\rm d}N_{\Delta_{\tilde{\tau}_{\tilde{1}}}}}{{\rm d}z_1}&=&0\,,
\end{eqnarray}
where $\varepsilon_{1\tilde{\tau}_1}\equiv \varepsilon_{1e}+\varepsilon_{1\mu}$. Thus the frozen values of the asymmetries at the end of $N_1$ washout are given by
\begin{eqnarray}
\nonumber N_{\Delta_\tau}(T\simeq T_{B_1})&=&\dfrac{3}{4}\varepsilon_{1\tau}\kappa(K_{1\tau})+N_{\Delta_\tau}(T\simeq T_\star)\,e^{-\frac{3\pi}{8}K_{1\tau}} \,, \\
\nonumber N_{\Delta_{\tilde{\tau}_1}}(T\simeq T_{B_1})&=&\dfrac{3}{4}\varepsilon_{1\tilde{\tau}_1}\kappa(K_{1\tilde{\tau}_1})+N_{\Delta_{\tilde{\tau}_1}}({T\simeq T_\star})\,e^{-\frac{3\pi}{8}K_{1\tilde{\tau}_1}} \,, \\
N_{\Delta_{\tilde{\tau}_{\tilde{1}}}}(T\simeq T_{B_1})&=&N_{\Delta_{\tilde{\tau}_{\tilde{1}}}}(T\simeq T_\star)
\end{eqnarray}
where $K_{1\tilde{\tau}_1}\equiv K_{1e}+K_{1\mu}$. Notice that decomposing $N_{\Delta \tilde{\tau}_2}(T\simeq T_\star)$ and $N_{\Delta \tilde{\tau}_{\tilde{2}}}(T\simeq T_\star)$ along the directions of $|\tilde{\tau}_1\rangle$ and $|\tilde{\tau}_{\tilde{1}}\rangle$ gives rise to
$N_{\Delta_{\tilde{\tau}_1}}({T\simeq T_\star})$ and $N_{\Delta_{\tilde{\tau}_{\tilde{1}}}}(T\simeq T_\star)$ as follows,
\begin{eqnarray}
\nonumber N_{\Delta_{\tilde{\tau}_1}}({T\simeq T_\star})&=&p_{\tilde{\tau}_2\tilde{\tau}_1}N_{\Delta_{\tilde{\tau}_2}}(T\simeq T_\star) + p_{\tilde{\tau}_{\tilde{2}}\tilde{\tau}_1}N_{\Delta_{\tilde{\tau}_{\tilde{2}}}}(T\simeq T_\star) \,, \\
N_{\Delta_{\tilde{\tau}_{\tilde{1}}}}(T\simeq T_\star)&=&(1-p_{\tilde{\tau}_2\tilde{\tau}_1})N_{\Delta_{\tilde{\tau}_2}}(T\simeq T_\star) + (1-p_{\tilde{\tau}_{\tilde{2}}\tilde{\tau}_1})N_{\Delta_{\tilde{\tau}_{\tilde{2}}}}(T\simeq T_\star) \,,
\end{eqnarray}
with
\begin{eqnarray}
\nonumber && p_{\tilde{\tau}_2\tilde{\tau}_1}=|\langle\tilde{\tau}_1|\tilde{\tau}_2\rangle|^2=\dfrac{\left|C^*_{1\mu}C_{2\mu}+C^*_{1e}C_{2e}\right|^2}{(1-p_{2\tau})(1-p_{1\tau})} \,, \\
&&p_{\tilde{\tau}_{\tilde{2}}\tilde{\tau}_1}=|\langle\tilde{\tau}_1|\tilde{\tau}_{\tilde{2}}\rangle|^2=\dfrac{\left|C^*_{1\mu}C_{3\mu}+C^*_{1e}C_{3e}-(C^*_{1\mu}C_{2\mu}+C^*_{1e}C_{2e})\sum_\alpha C^*_{2\alpha}C_{3\alpha}\right|^2}{(1-p_{1\tau})(1-p_{\tilde{2}\tau})(1-p_{32})} \,.
\end{eqnarray}
We would like to mention that the factors $C_{1\alpha}$, $C_{2\alpha}$, $C_{3\alpha}$ depend on the neutrino Yukawa, as shown in Eq.~\eqref{eq:C-par}. The total final asymmetry is the sum of the three terms
\begin{eqnarray}
N_{B-L}(T\simeq T_{B_1})=N_{\Delta_\tau}(T\simeq T_{B_1})+N_{\Delta_{\tilde{\tau}_1}}(T\simeq T_{B_1})+N_{\Delta_{\tilde{\tau}_{\tilde{1}}}}(T\simeq T_{B_1})\,.
\end{eqnarray}
It turns out that our benchmark models are in the strong washout regime with $K_3,K_2,K_{1\tau},K_{1\tilde{\tau}_1}\gg1$, thus we neglect the contributions which undergo the washout exponential suppression of at least one right-handed neutrino. The final lepton asymmetry will be then dominated by the unwashed terms and it is approximately given by
\begin{eqnarray}
\nonumber N_{B-L}^{\rm f}&=&\dfrac{3}{4}\Big[\varepsilon_{1\tau}\kappa(K_{1\tau}) + \varepsilon_{1\tilde{\tau}_1}\kappa(K_{1\tilde{\tau}}) + (1-p_{\tilde{\tau}_2\tilde{\tau}_1})(1-p_{2\tau})\varepsilon_2\kappa(K_2)  \\
&&~~~+(1-p_{\tilde{\tau}_{\tilde{2}}\tilde{\tau}_1})(1-p_{\tilde{2}\tau})(1-p_{32})\varepsilon_3\kappa(K_3)\Big]\,.
\end{eqnarray}
Notice that this approximation formula is applicable to both thermal and vanishing initial abundance of right-handed neutrinos, and the corresponding efficiency factors are given in Eq.~\eqref{eq:kappa-thermal} and Eq.~\eqref{eq:kappa-vanishing} respectively.

\section{Numerical results}
\label{section4}

\begin{table}[t!]
\centering
\begin{tabular}{|c|c|c|c|} \hline  \hline
Parameters & $\mu_i\pm1\sigma$ & Parameters & $\mu_i\pm1\sigma$ \\ \hline
$m_{t}/\text{GeV}$ & $83.155 \pm 3.465$ & $\theta_{12}^{q}$ & $0.229 \pm 0.001$ \\
$m_{b}/\text{GeV}$ & $0.884 \pm 0.035$ & $\theta_{13}^{q}$ & $0.0037 \pm 0.0004$ \\
  $m_{u}/m_{c}$ & $0.0027 \pm 0.0006$ & $\theta_{23}^{q}$ & $0.0397 \pm 0.0011 $ \\
  $m_{c}/m_{t}$ & $0.0025 \pm 0.0002$ & $\delta_{CP}^{q}/^{\circ}$ & $56.34 \pm 7.89 $ \\
  $m_{d}/m_{s}$ & $0.051 \pm 0.007$ & $\sin^{2}\theta_{12}^{l}$ & $0.304\pm 0.012$ \\
  $m_{s}/m_{b}$ & $0.019 \pm 0.002$ & $\sin^{2}\theta_{23}^{l}$ & $0.450_{-0.016}^{+0.019}$ \\
  $m_{e}/m_{\mu}$ & $0.0048 \pm 0.0002$ & $\sin^{2}\theta_{13}^{l}$ & $0.02246\pm 0.00062$ \\
  $m_{\mu}/m_{\tau}$ & $0.059 \pm 0.002$ & $\delta_{CP}^{l}/^{\circ}$ & $230_{-25}^{+36}$ \\
$m_{b}/m_{\tau}$ & $0.73 \pm 0.03$ & $r\equiv\Delta m_{21}^{2}/\Delta m_{31}^{2}$ & $0.02956 \pm 0.00084$\\
$Y_B$ & $(8.72\pm 0.27)\times 10^{-11}$ & $\Delta m_{21}^{2}/\text{eV}^{2}$ & $7.42^{+0.21}_{-0.20}\times 10^{-5}$\\
\hline \hline
\end{tabular}
\caption{\label{tab:parameter-values-GUT} The best fit values $\mu_i$ and $1\sigma$ uncertainties of the quark and lepton parameters when evolved to the GUT scale as calculated in~\cite{Ross:2007az}, with the SUSY breaking scale $M_{\text{SUSY}}=500$ GeV and $\tan\beta=10$, where the error widths represent $1\sigma$ intervals. The values of lepton mixing angles, leptonic Dirac CP violation phases $\delta^{l}_{CP}$ and the neutrino mass squared difference are taken from NuFIT 5.1~\cite{Esteban:2020cvm} for normal ordering neutrino masses with Super-Kamiokande data. }
\end{table}

In this section, we shall perform a detailed numerical analysis to explore whether the renormalizable $SO(10)\times A_4$ modular models of section~\ref{subsec:benchmark-models} can explain the experimentally measured
values of masses and mixing parameters of both quarks and leptons as well as the matter-antimatter asymmetry of the Universe. We perform a $\chi^2$ analysis to optimize the values of the free parameters.  The $\chi^{2}$ function is defined as
\begin{equation}
\chi^2=\sum_{i}\left(\frac{P_i(x)-\mu_i}{\sigma_i}\right)^2\,.
\end{equation}
Here $\mu_i$ and $\sigma_i$ denote the experimental central values and the $1\sigma$ uncertainties of the observables respectively, and their values are obtained by evolving their low energy values to the GUT scale with the renormalization group equations, as shown in table~\ref{tab:parameter-values-GUT}. For the neutrino oscillation data, we assume normal ordering neutrino mass spectrum so that the renormalization group induced corrections to the neutrino masses and mixing parameters
can be negligible for small $\tan\beta$, and we use the most recent results from the NuFit collaboration~\cite{Esteban:2020cvm}.

The mechanism of modulus stabilization is still an open question, consequently we treat the complex modulus $\tau$ as a free parameter in the fundamental region $\mathcal{D}=\left\{\tau\in\mathbb{C}\Big| \text{Im}\tau>0, |\text{Re}\tau|\leq\frac{1}{2}, |\tau|\geq1\right\}$, which can represent all possible values of $\tau$ in the upper half complex plane up to a modular transformation. In the numerical analysis, the magnitudes of all coupling constants are limited in the region $[0, 10^4]$ and their phases are freely varied in the range $[0, 2\pi]$. The overall scales $\alpha_1 v_u$, $\alpha_1 r_1v_u$ and $\alpha^2_1v^2_u/v_R$ are fixed by the experimentally measured values of top quark mass, bottom quark mass and solar neutrino mass squared difference $\Delta m^2_{21}$. For each set of given values of the input parameters, the quark and lepton mass matrices in Eqs.~(\ref{eq:mass-matrix-charged-fermions}, \ref{eq:nu-matrix}) can be straightforwardly diagonalized and subsequently mixing angles and CP violation phases can be extracted in the usual way.
We numerically minimize the  $\chi^2$ function by using the minimization algorithms TMinuit~\cite{minuit} developed by CERN to determine the optimum values of the input parameters. In the present work, we will focus the scenario of type I seesaw dominance, and the out of equilibrium decay of right-handed neutrinos is used to generate the baryon asymmetry of the Universe.

For each model, we present the point in parameter space which
minimizes the $\chi^2$, as a result of a numerical minimization procedure. We shall consider two typical initial conditions of heavy Majorana neutrinos:
thermal initial abundance and vanishing initial abundance. We show the numerical results in table~\ref{tab:fit-non_minimal_model_2-VIA} and table~\ref{tab:fit-non_minimal_model_2_second-TIA} for vanishing and thermal initial conditions respectively. It is notable that the value of $\tau$ is close to the residual modular symmetry fixed point $\tau_{ST}\equiv-1/2+i\sqrt{3}/2$. Since both models at the best points are in the strong washout regime, the prediction for the baryon asymmetry is not so sensitive to the initial condition of right-handed neutrinos. Thus the values of each input parameter have little difference for the concerned two initial conditions. We also present the predictions for the mass ratios of charged fermions $m_e/m_{\mu}$, $m_{\mu}/m_{\tau}$, $m_u/m_c$, $m_c/m_t$, $m_d/m_s$, $m_s/m_b$ and $m_b/m_{\tau}$, the mixing angles $\theta^{l}_{12}$, $\theta^{l}_{13}$, $\theta^{l}_{23}$, $\theta^{q}_{12}$, $\theta^{q}_{13}$ and $\theta^{q}_{23}$ as well as the quark CP violation phase $\delta^q_{CP}$, which are compatible with experimental data. Moreover, from the fitted values of the parameters, we can derive the predictions for the still unmeasured observables including the leptonic Dirac CP violation phase $\delta^{l}_{CP}$ and Majorana CP phases $\alpha_{21}$ and $\alpha_{31}$, the light neutrino masses $m_{1,2,3}$, the effective Majorana neutrino mass $m_{\beta\beta}$ for neutrinoless double decay and the right-handed neutrino masses $M_{1,2,3}$ at the best fit point. The right-handed neutrino masses have a hierarchical pattern $M_3\gg M_2\gg M_1$ which are in the range of $10^{10}-10^{13}$ GeV. The effective Majorana mass $m_{\beta\beta}$ is determined to be $0.532$ (1.059) meV and 0.622 (0.550) meV in the models \texttt{BM1} and \texttt{BM2} respectively for vanishing (thermal) initial condition, they are much below the current most stringent bound $m_{\beta\beta}<(61-165)$ meV given by the KamLAND-Zen Collaboration~\cite{KamLAND-Zen:2016pfg}. Future tonne-scale neutrinless double decay experiments can reduce the sensitivity on $m_{\beta\beta}$ to few meV such as $m_{\beta\beta}<(4.7-20.3)$ meV from nEXO~\cite{nEXO:2021ujk} and $m_{\beta\beta}<(9-21)$ meV from LEGEND-1000~\cite{LEGEND:2021bnm}. Consequently our predictions for $m_{\beta\beta}$ are even out of reach of tonne-scale detectors. As shown in tables~\ref{tab:fit-non_minimal_model_2-VIA} and~\ref{tab:fit-non_minimal_model_2_second-TIA}, we have split the $\chi^2$ function into four different contributions $\chi^2_{total}=\chi^2_{l}+\chi^2_q+\chi^2_{b\tau}+\chi^2_{Y_B}$, where $\chi^2_l$, $\chi^2_q$, $\chi^2_{b\tau}$ and $\chi^2_{Y_B}$ stand for the contributions induced by the deviations of the lepton sector observables, the
quark sector observables, the mass ratio $m_b/m_{\tau}$ and the baryon asymmetry $Y_B$ from their central values respectively. We see that the largest deviation arises from the quark sector in both models, and $\chi^2_{total}$ is dominated by $\chi^2_q$ although the predictions for the quark masses and mixing parameters are in the experimentally allowed region.

\begin{figure}[hptb!]
\includegraphics[width=6.5in]{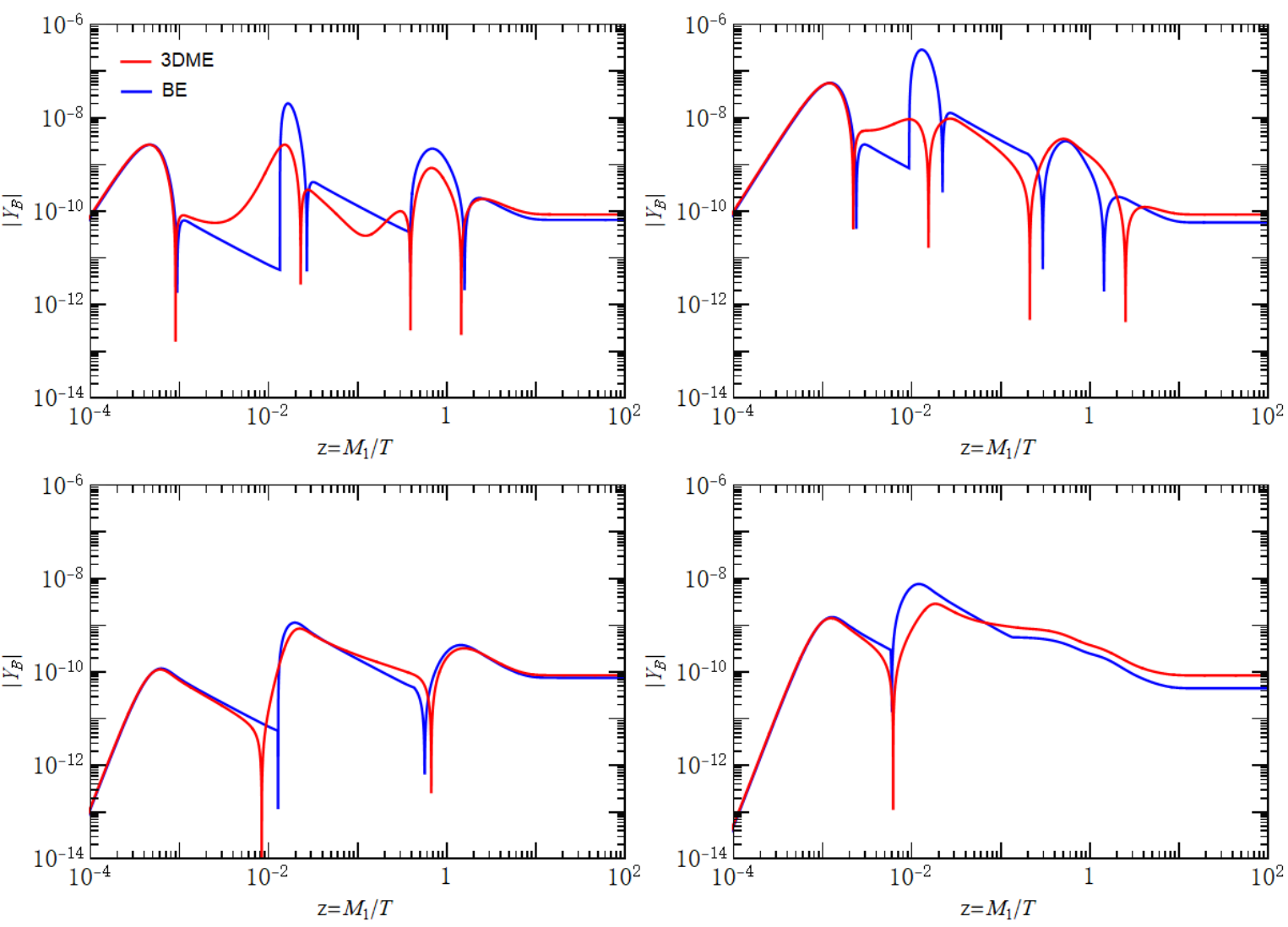}
\caption{\label{fig:Yb-T-evolution}The evolution of the baryon asymmetry $Y_B$ with temperature for the two benchmark models, the free parameters take the best fit values in table~\ref{tab:fit-non_minimal_model_2-VIA} and table~\ref{tab:fit-non_minimal_model_2_second-TIA}. The red line and blue line are the numerical results obtained by solving the density matrix equation and Boltzmann equation of section~\ref{subsec:analytical-app} respectively. The top-left\,(right) panel is for the $\texttt{BM1}\,(\texttt{BM2})$ model with vanishing initial abundance. The bottom-left (right) panel for the \texttt{BM1}\,(\texttt{BM2}) model with thermal initial abundance.}
\end{figure}

Regarding the baryon asymmetry, the concrete form of the neutrino Yukawa coupling matrix and right-handed neutrino masses are fixed at the best fit point, we numerically solve the density matrix equation of Eq.~\eqref{eq:dme} with the help of ULYSSES to obtain the frozen value of $B-L$ asymmetry. It is remarkable that the numerical results agree well with the analytical approximations. We see that the observed matter-antimatter asymmetry can be accommodated. In general, the decays of three right-handed neutrinos $N_{1,2,3}$ all contribute to the final baryon asymmetry, and asymmetries generated by $N_{2,3}$ are partially washed out by the $N_1$ process. In the model \texttt{BM1} for both initial conditions and the \texttt{BM2} model with vanishing initial abundance, the leptogensis is dominated by $N_1$ decay. The contributions of $N_2$ and $N_3$ are as important as the that of $N_1$ in the \texttt{BM2} for thermal initial abundance. Moreover, we plot the evolution of $Y_B$ with the temperature $T$ for the two benchmark models in figure~\ref{fig:Yb-T-evolution}, and the corresponding numerical results of Boltzmann equations are displayed for comparison.

Furthermore, we comprehensively scan the parameter space around the best fit points of table~\ref{tab:fit-non_minimal_model_2-VIA} and table~\ref{tab:fit-non_minimal_model_2_second-TIA}, and we require all the fermion masses and mixing parameters are in the experimentally preferred $3\sigma$ regions. The experimentally allowed values of the complex modulus $\tau$ and the correlations between the masses and mixing parameters are shown in figures~\ref{fig:non_minimal_2-VIA}, \ref{fig:non_minimal_3-VIA}, \ref{fig:non_minimal_2-TIA} and \ref{fig:non_minimal_3-TIA} for the benchmark models with vanishing and thermal initial conditions. The corresponding predictions for the baryon asymmetry $Y_B$ with respect to $\sin^2\theta^l_{23}$, $\delta^{l}_{CP}$, $\alpha_{21}$, $\alpha_{31}$ are displayed in figures~\ref{fig:non_minimal_2-YB-VIA}, \ref{fig:non_minimal_3-YB-VIA}, \ref{fig:non_minimal_2-YB-TIA} and \ref{fig:non_minimal_3-YB-TIA}. We would like to mention that the contribution of $Y_B$ to $\chi^2$ is not included in these figures. We see that imposing successful leptogenesis yields specific predictions for  the fermion masses and mixing parameters.

\begin{table}[]
 \centering
  \begin{tabular}{|c|c|c|}
\hline \hline
 \texttt{parameters} &   \texttt{BM1} &  \texttt{BM2} \\ \hline
$\tau$ & $-0.4982+0.8862 i$ & $-0.4796+0.8958 i$\\
$r_2$ & $1.293~e^{i 1.790\pi}$ & $1.572~e^{i 1.219\pi}$\\
$r_3$ & $0.343~e^{i 0.395\pi}$ & $1.374~e^{i 1.921\pi}$\\
$c_e$ & $9.891~e^{i 1.447\pi}$ & $1.591~e^{i 0.731\pi}$\\
$c_\nu$ & $7.485~e^{i 0.108\pi}$ & $0.952~e^{i 1.644\pi}$\\
$\alpha_{2}/\alpha_{1}$ & $0.310~e^{i 1.839\pi}$ & $0.600~e^{i 1.025\pi}$\\
$\alpha_{3}/\alpha_{1}$ & $0.307~e^{i 1.838\pi}$ & $0.597~e^{i 1.034\pi}$\\
$\beta_{1}/\alpha_{1}$ & $0.093~e^{i 0.862\pi}$ & $3.481~e^{i 0.139\pi}$\\
$\beta_{2}/\alpha_{1}$ & --- & $0.025~e^{i 1.606\pi}$\\
$\gamma_{1}/\alpha_{1}$ & $0.295~e^{i 1.400\pi}$ & $0.243~e^{i 0.664\pi}$\\
$\gamma_{2}/\alpha_{1}$ & $0.297~e^{i 0.094\pi}$ & $0.112~e^{i 1.550\pi}$\\
$\gamma_{3}/\alpha_{1}$ & $0.300~e^{i 0.095\pi}$ & $0.114~e^{i 1.555\pi}$\\
$\alpha_1^2 v_u^2/v_R({\rm meV})$ & $15.272$ & $7.053$\\
$\alpha_1 v_u/{\rm GeV}$ & $20.673$ & $18.157$\\
$\alpha_1 r_1 v_d$/{\rm GeV} & $0.272$ & $0.240$\\
\hline \hline
$\sin^2\theta_{12}^l$ & $0.308$ & $0.294$\\
$\sin^2\theta_{13}^l$ & $0.02178$ & $0.02248$\\
$\sin^2\theta_{23}^l$ & $0.465$ & $0.455$\\
$\delta_{CP}^l/^\circ$ & $267.074$ & $333.240$\\
$\alpha_{21}/^\circ$ & $212.233$ & $222.547$\\
$\alpha_{31}/^\circ$ & $237.998$ & $31.091$\\
\hline $m_e/m_\mu$ & $0.00488$ & $0.00484$\\
$m_\mu/m_\tau$ & $0.0584$ & $0.0586$\\
$m_1/{\rm meV}$ & $3.065$ & $2.645$\\
$m_2/{\rm meV}$ & $9.143$ & $9.011$\\
$m_3/{\rm meV}$ & $50.597$ & $50.375$\\
$m_{\beta\beta}/{\rm meV}$ & $0.532$ & $0.622$\\
$M_1/{\rm GeV}$ & $6.546\times 10^{10}$ & $7.447\times 10^{10}$\\
$M_2/{\rm GeV}$ & $1.716\times 10^{12}$ & $3.307\times 10^{12}$\\
$M_3/{\rm GeV}$ & $4.853\times 10^{13}$ & $3.012\times 10^{13}$\\
$v_R/{\rm GeV}$ & $2.798\times 10^{13}$ & $4.674\times 10^{13}$\\
\hline $\theta_{12}^q$ & $0.230$ & $0.229$\\
$\theta_{13}^q$ & $0.00411$ & $0.00332$\\
$\theta_{23}^q$ & $0.0432$ & $0.0421$\\
$\delta_{CP}^q/^\circ$ & $83.606$ & $76.967$\\
\hline $m_u/m_c$ & $0.00329$ & $0.00285$\\
$m_c/m_t$ & $0.00286$ & $0.00287$\\
$m_d/m_s$ & $0.0353$ & $0.0491$\\
$m_s/m_b$ & $0.0169$ & $0.0206$\\
\hline $m_b/m_\tau$ & $0.715$ & $0.803$\\
\hline $Y_{B}^{\rm num}$ & $8.541 \times 10^{-11}$ & $8.516 \times 10^{-11}$\\
$Y_B^{\rm ap}$ & $6.806\times 10^{-11}$ & $6.355\times 10^{-11}$ \\
\hline $\chi^2_l$ & $3.523$ & $9.202$\\
$\chi^2_q$ & $33.912$ & $16.555$\\
$\chi^2_{b\tau}$ & $0.264$ & $5.958$\\
$\chi^2_{Y_{B}}$ & $0.441$ & $0.573$\\
$\chi^2_{total}$ & $38.140$ & $31.715$\\
\hline \hline
\end{tabular}
\caption{\label{tab:fit-non_minimal_model_2-VIA}The best fit values of the free parameters and the corresponding predictions for the masses and mixing parameters of leptons and quarks as well as the baryon asymmetry $Y_B$ in the two typical models \texttt{BM1} and \texttt{BM2}. We assume vanishing initial abundance of the right-handed neutrinos, $Y_{B}^{\rm num}$ is the numerical result from density matrix equation, and $Y_{B}^{\rm ap}$ denotes the result of analytical approximation in section~\ref{subsec:analytical-app}. }
\end{table}

\begin{table}[]
\centering
\begin{tabular}{|c|c|c|}
\hline\hline
\texttt{parameters} &   \texttt{BM1} &  \texttt{BM2}  \\ \hline
$\tau$ & $-0.4949+0.8861 i$ & $-0.4997+0.8975 i$\\
$r_2$ & $1.414~e^{i 1.811\pi}$ & $2.441~e^{i 1.377\pi}$\\
$r_3$ & $0.315~e^{i 0.268\pi}$ & $1.910~e^{i 1.903\pi}$\\
$c_e$ & $10.425~e^{i 1.365\pi}$ & $1.675~e^{i 0.613\pi}$\\
$c_\nu$ & $7.276~e^{i 0.149\pi}$ & $0.890~e^{i 1.214\pi}$\\
$\alpha_{2}/\alpha_{1}$ & $0.286~e^{i 1.912\pi}$ & $0.626~e^{i 1.011\pi}$\\
$\alpha_{3}/\alpha_{1}$ & $0.283~e^{i 1.912\pi}$ & $0.631~e^{i 1.016\pi}$\\
$\beta_{1}/\alpha_{1}$ & $0.104~e^{i 0.774\pi}$ & $3.590~e^{i 0.104\pi}$\\
$\beta_{2}/\alpha_{1}$ & --- & $0.016~e^{i 1.247\pi}$\\
$\gamma_{1}/\alpha_{1}$ & $0.197~e^{i 1.405\pi}$ & $0.333~e^{i 0.639\pi}$\\
$\gamma_{2}/\alpha_{1}$ & $0.311~e^{i 0.102\pi}$ & $0.140~e^{i 1.600\pi}$\\
$\gamma_{3}/\alpha_{1}$ & $0.313~e^{i 0.102\pi}$ & $0.141~e^{i 1.597\pi}$\\
$\alpha_1^2 v_u^2/v_R({\rm meV})$ & $12.574$ & $8.865$\\
$\alpha_1 v_u/{\rm GeV}$ & $19.718$ & $13.894$\\
$\alpha_1 r_1 v_d$/{\rm GeV} & $0.266$ & $0.242$\\
\hline \hline
$\sin^2\theta_{12}^l$ & $0.321$ & $0.305$\\
$\sin^2\theta_{13}^l$ & $0.02258$ & $0.02235$\\
$\sin^2\theta_{23}^l$ & $0.438$ & $0.433$\\
$\delta_{CP}^l/^\circ$ & $329.478$ & $229.431$\\
$\alpha_{21}/^\circ$ & $175.644$ & $184.054$\\
$\alpha_{31}/^\circ$ & $240.818$ & $287.703$\\
\hline $m_e/m_\mu$ & $0.00481$ & $0.00502$\\
$m_\mu/m_\tau$ & $0.0585$ & $0.0599$\\
$m_1/{\rm meV}$ & $2.156$ & $7.053$\\
$m_2/{\rm meV}$ & $8.880$ & $11.133$\\
$m_3/{\rm meV}$ & $49.719$ & $50.466$\\
$m_{\beta\beta}/{\rm meV}$ & $1.059$ & $0.550$\\
$M_1/{\rm GeV}$ & $7.083\times 10^{10}$ & $2.798\times 10^{10}$\\
$M_2/{\rm GeV}$ & $1.682\times 10^{12}$ & $1.793\times 10^{12}$\\
$M_3/{\rm GeV}$ & $5.608\times 10^{13}$ & $1.739\times 10^{13}$\\
$v_R/{\rm GeV}$ & $3.092\times 10^{13}$ & $2.177\times 10^{13}$\\
\hline $\theta_{12}^q$ & $0.229$ & $0.230$\\
$\theta_{13}^q$ & $0.00359$ & $0.00416$\\
$\theta_{23}^q$ & $0.0421$ & $0.0440$\\
$\delta_{CP}^q/^\circ$ & $55.549$ & $86.564$\\
\hline $m_u/m_c$ & $0.00278$ & $0.00285$\\
$m_c/m_t$ & $0.00274$ & $0.00260$\\
$m_d/m_s$ & $0.0302$ & $0.0446$\\
$m_s/m_b$ & $0.0173$ & $0.0156$\\
\hline $m_b/m_\tau$ & $0.708$ & $0.702$\\
\hline $Y_{B}^{\rm num}$ & $8.526\times 10^{-11}$ & $8.458\times 10^{-11}$\\
$Y_B^{\rm ap}$ & $7.800\times 10^{-11}$ & $5.215\times 10^{-11}$ \\
\hline $\chi^2_l$ & $10.694$ & $2.539$\\
$\chi^2_q$ & $16.122$ & $35.517$\\
$\chi^2_{b\tau}$ & $0.554$ & $0.847$\\
$\chi^2_{Y_{B}}$ & $0.518$ & $0.940$\\
$\chi^2_{total}$ & $27.888$ & $38.843$\\
\hline \hline
\end{tabular}
\caption{\label{tab:fit-non_minimal_model_2_second-TIA}
Similar to table~\ref{tab:fit-non_minimal_model_2-VIA}, but for thermal initial abundance of the right-handed neutrinos. }
\end{table}

\section{\label{sec:conclusion}Conclusion}

In this paper we have studied the prediction for leptogenesis in two renormalizable supersymmetric $SO(10)\times A_4$ modular models in which the neutrino mass is dominantly generated by the type I seesaw mechanism. The evolution of the lepton asymmetries are described in terms of the three-flavored density matrix equations for three heavy Majorana neutrinos, where both vanishing initial condition and thermal initial condition of the right-handed neutrinos are considered. We also presented an analytical approximation based on the Boltzmann equations. We found regions of parameter space compatible with the measured fermion masses and mixing parameters as well as the baryon asymmetry of the Universe. The predictions for the light neutrino masses, the effective mass in neutrinoless doble beta decay and the leptonic CP violation phases were discussed.

The right-handed neutrinos naturally appear in $SO(10)$ GUT, all the chiral fermions of one generation plus one additional right-handed neutrino are unified into a single 16 dimensional spinor representation of $SO(10)$. Thus $SO(10)$ GUT provides a natural explanation for the tiny neutrino mass through the type I seesaw mechanism. From the view of cosmology, it is well known that the CP violating decay of the right-handed neutrinos can generate the matter-antimatter asymmetry of the Universe, this is the so-called leptogenesis. Moreover, the mass and Yukawa coupling of the right-handed neutrinos are closely related to those of quarks and charged leptons, and their concrete values can be determined by requirement that the measured masses, mixing angles and CP violation phases of both quarks and leptons should be accommodated. It is attractive that the fermion masses and mixing together with baryon asymmetry of the Universe can be explained in a given $SO(10)$ model.

Modular symmetry is an important progress to address the flavor puzzle of the SM, the vacuum alignment problem in conventional flavor symmetry models is greatly simplified, and the complex modulus $\tau$ could be the unique source of flavor symmetry breaking. As a consequence, the modular symmetry models could be quite predictive. In the present work, we have analyzed the prediction for leptogenesis in two benchmark renormalizable $SO(10)$ models with $A_4$ modular symmetry where the neutrino masses dominantly arise from type I seesaw mechanism. The three generations of matter fields are assigned to triplet of $A_4$ modular symmetry, and the Higgs fields $H$,  $\Sigma$, $\overline{\Delta}$ in the 10, 120 and 16 dimensional representations of $SO(10)$ are invariant under $A_4$. As a consequence, the structure of the $SO(10)\times A_4$ modular models are fully specified by the modular weights of the fermion fields and Higgs fields, as shown in table~\ref{tab:SO10-A4-modular-model}. The modular weights are  $(2k_F+k_{10}\,, 2k_F+k_{120}\,, 2k_F+k_{\overline{126}})=(4,2,4)$ and $(2k_F+k_{10}\,, 2k_F+k_{120}\,, 2k_F+k_{\overline{126}}) = (4,6,4)$ respectively in the two concerned benchmark models.

We used the density matrix equations to describe the kinetic evolution of the lepton asymmetry. In comparison with Boltzmann equations, the density matrix equations give a quantitative description of lepton flavor effects in leptogenesis, and allows to precisely calaulate the final lepton asymmetry for arbitrary right-handed neutrino mass spectrum. The right-handed neutrino masses are hierarchical with $M_3\gg M_2\gtrsim10^{12}\,{\rm GeV}$ and $10^9\,{\rm GeV}\ll M_1\ll 10^{12}\,{\rm GeV}$ in our benchmark models. Hence the decays of $N_3$ and $N_2$ are in the single-flavored regime while $N_1$ decay is in the two-flavored regime. In order to understand the physical process of leptogenesis, we assume the three heavy Majorana neutrinos decay in sequence. The lepton asymmetries generated from the decay of each right-handed neutrino and the washout effects are governed by the Boltzmann equation, the analytical approximate solutions are given. The approximation agrees with the numerical results of density matrix equations well if the interplay of the different right-handed neutrinos and the interference of lepton flavor are insignificant.

The superpotential of the Yukawa is strongly constrained by modular symmetry, it involves 7 couplings $\alpha_{1, 2, 3}$, $\beta_{1}$ and $\gamma_{1, 2, 3}$ in the \texttt{BM1} model, and an additional parameter $\beta_2$ in the \texttt{BM2} model. Moreover, the quark and lepton mass matrices depend on 5 parameters $r_{1, 2, 3}$, $c_{e, \nu}$ which describe the light Higgs combinations. Notice that the Higgs sector of the renormalizable $SO(10)$ models with all  10, 120 and 126 dimensional Higgs multiplets are complicated and it reduces the predictive power of $SO(10)$ models. The 126-plet Higgs $\overline{\Delta}$ contains a right-handed scalar triplet $\Delta_R$ whose vacuum expectation value $\langle\Delta_R\rangle=v_R$ gives mass to the right-handed neutrinos. The parameters $\alpha_1$, $r_1$ and $v_R$ can be taken real without loss of generality. Including the complex modulus $\tau$, the \texttt{BM1} and \texttt{BM2} models depend on 25 and 27 real parameters respectively. We fit all these free parameters using the 20 observables in table~\ref{tab:parameter-values-GUT}. In particular, we take into account the  matter-antimatter asymmetry of the Universe through leptogenesis besides the measured masses and mixing parameters of quarks and leptons. As shown in table~\ref{tab:fit-non_minimal_model_2-VIA} and table~\ref{tab:fit-non_minimal_model_2_second-TIA}, the experimental data can be well accommodated and it is remarkable that all the coupling constants are order one. Furthermore, we can obtain predictions for unmeasured quantities such as the Majorana CP phases $\alpha_{21}$, $\alpha_{31}$, the effective mass $m_{\beta\beta}$ of neutrinoless double decay and the right-handed neutrino masses $M_{1, 2, 3}$. $m_{\beta\beta}$ is predicted in the range 0.5-1.1 meV, it is far below the sensitivity of current and future experiments. Hence our models would be ruled out if a positive signal of neutrinoless double beta decay is reported in future.

\section*{Acknowledgements}

GJD, JNL and BYQ are supported by the National Natural Science Foundation of China under Grant Nos. 11975224, 11835013 and the Key Research Program of the Chinese Academy of Sciences under Grant NO. XDPB15. SFK acknowledges the STFC Consolidated Grant ST/L000296/1 and the European Union's Horizon 2020 Research and Innovation programme under Marie Sk\l{}odowska-Curie grant agreement HIDDeN European ITN project (H2020-MSCA-ITN-2019//860881-HIDDeN).

\section*{Appendix}

\setcounter{equation}{0}
\renewcommand{\theequation}{\thesection.\arabic{equation}}

\begin{appendix}

\section{\label{sec:app-bad-examples} Two examples of analytical approximation failure }

The formalism of density matrix allows for a more general description of leptogenesis than the semi-classical Boltzmann equations. The off-diagonal elements of the density matrix describe the degree of coherence between the flavour states, so that the decoherence effects are incorporated into the equations. Moreover, the density matrix equations can describe the dynamical process of the transition between flavour-regimes, and allows to calculate the lepton asymmetry in the intermediate regimes where the single-flavored and two-flavored treatments are inadequate. Hence it is found that the Boltzmann equations fail to describe correctly the generation of baryon asymmetry in many cases~\cite{Moffat:2018wke,Moffat:2018smo,Granelli:2021fyc}.  Although the analytical approximation agree well with the numerical results of the density matrix equation in section~\ref{section4}, we note that this not always true. In particular, when the effects of decoherence cannot be neglected or there are non-trivial interplay between the decays and the washout processes of the three right-handed neutrinos. We give two such benchmark points in table~\ref{tab:fit-non_minimal_models-fail} at which our analytical estimates fail to give the correct value of baryon asymmetry, and even the sign of $Y_B$ is wrong.

\begin{figure}[hptb!]
\includegraphics[width=6.5in]{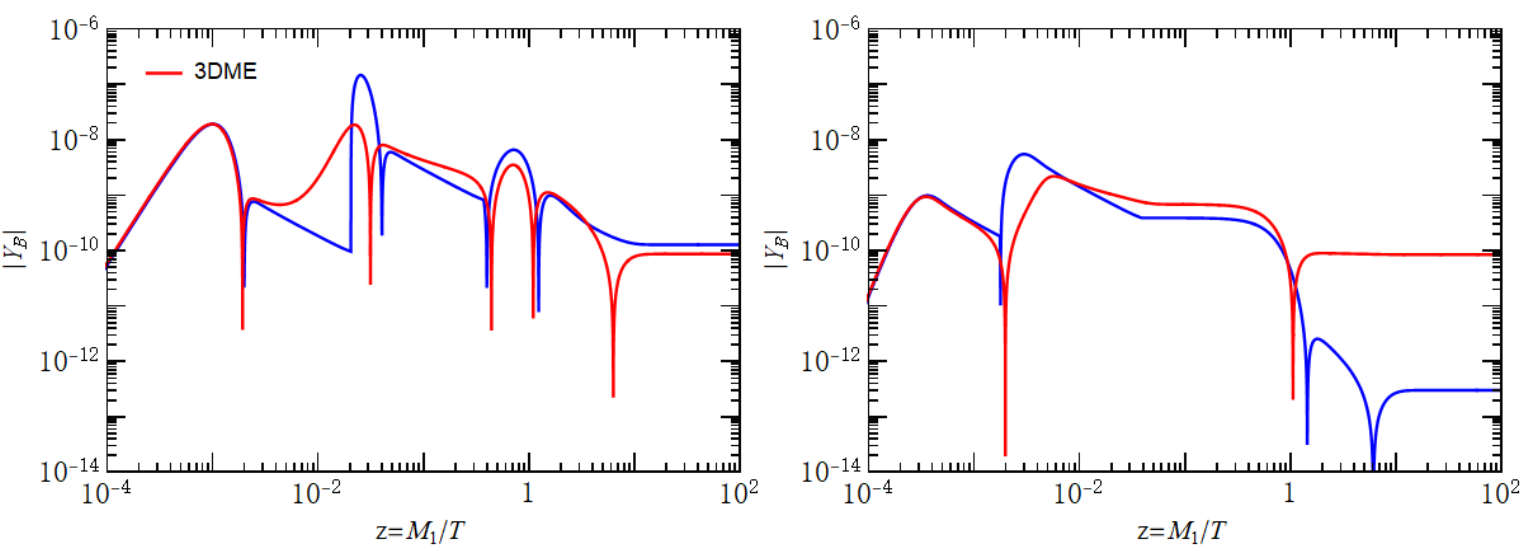}\\
\includegraphics[width=6.5in]{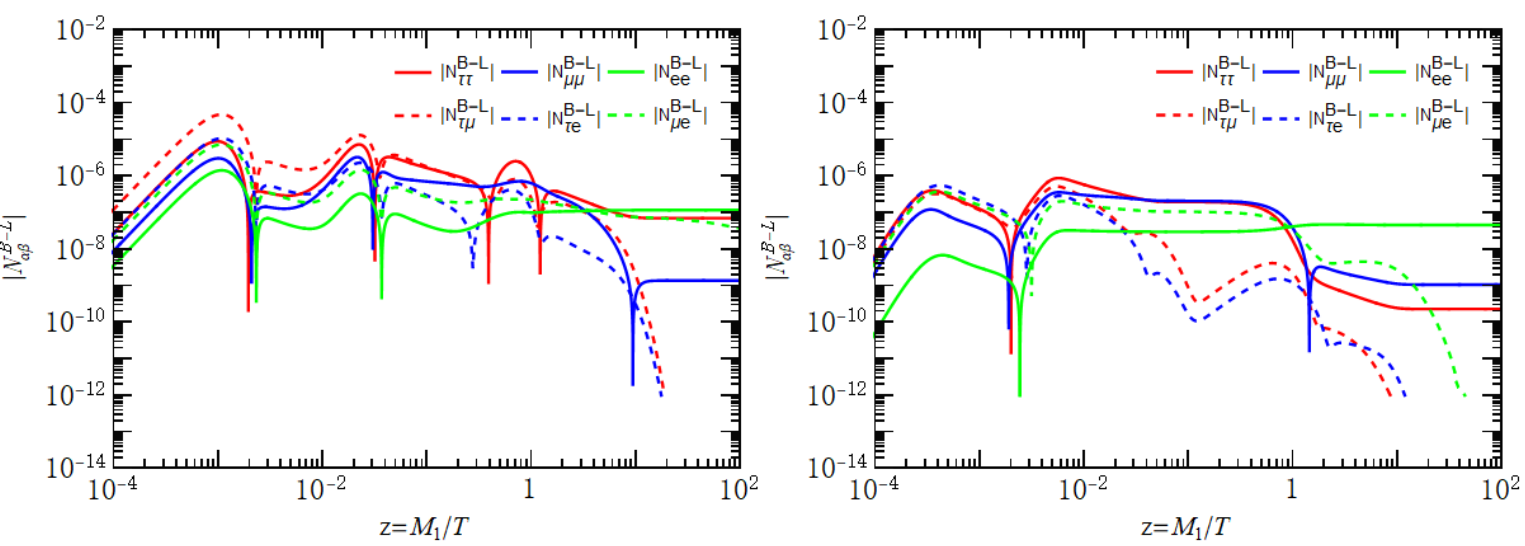}
\caption{\label{fig:Yb-T-evolution-fail}The evolution of the baryon asymmetry $Y_B$ with temperature for the  local minimum shown in table~\ref{tab:fit-non_minimal_models-fail}. The left and right panels are the \texttt{BM1} with vanishing initial abundance and the \texttt{BM2} with thermal initial abundance respectively. The evolution of the $B-L$ asymmetry number densities $N^{B-L}_{\alpha\beta}$, $\alpha,\beta=e, \mu, \tau$ displayed in the lower panel is described by the density matrix equations in Eq.~\eqref{eq:dme}.  }
\end{figure}

For the model \texttt{BM1} with the values of the free parameters given in table~\ref{tab:fit-non_minimal_models-fail}, the decay parameters of $N_2$ and $N_3$ are determined to be $K_2=90$ and $K_3=98$ respectively. Therefore the freeze-out temperature of $N_3$ is approximately $T_{B_3}\simeq M_3/9\simeq 2M_2$, the effects of $N_2$ are not negligible before the freeze-out of $N_3$ inverse decay, as can be seen from the left-top panel of figure~\ref{fig:Yb-T-evolution-fail}. In addition, the asymmetries generated by $N_2$ and $N_3$ are only partially washed out by the $N_1$ decay and inverse decay processes and in fact they dominate the final asymmetry. Moreover, the mass of $N_1$ is $2.308\times 10^{11}\,{\rm GeV}$, consequently it is in the transition region between the single- and two-flavored leptogenesis. This might contribute to the failure of the analytical estimates, since the transitions between the different flavour regimes are not taken into account in the Boltzmann equations and the baryon asymmetry $\eta$ could change its sign during the transitions~\cite{Granelli:2021fyc}. Analogously, for another model \texttt{BM2}, there is also non-trivial interplay between the decay and the washout processes of the heavy right-handed neutrinos. The contribution of $N_1$ is found to be subdominant, and the $N_1$ mass is $8.367\times 10^9$ GeV which is close to the transition scale between the two-flavor and three-flavor regimes. Furthermore, from figure~\ref{fig:Yb-T-evolution-fail} we see that the decoherence of flavor states is not infinitely fast as assumed in the Boltzmann equations. This can also lead to deviations of analytical approximation from the numerical results of density matrix.


\begin{table}[]
 \centering
  \begin{tabular}{|c|c|c|} \hline \hline
\texttt{parameters} &  \texttt{BM1}  & \texttt{BM2} \\ \hline

$\tau$ & $-0.4863+0.8863 i$ & $-0.5+0.9031 i$\\
$r_2$ & $1.305~e^{i 1.880\pi}$ & $2.067~e^{i 1.488\pi}$\\
$r_3$ & $0.267~e^{i 0.402\pi}$ & $1.731~e^{i 1.902\pi}$\\
$c_e$ & $8.188~e^{i 1.401\pi}$ & $1.247~e^{i 0.785\pi}$\\
$c_\nu$ & $7.473~e^{i 0.139\pi}$ & $0.890~e^{i 0.999\pi}$\\
$\alpha_{2}/\alpha_{1}$ & $0.286~e^{i 1.852\pi}$ & $0.555~e^{i 0.948\pi}$\\
$\alpha_{3}/\alpha_{1}$ & $0.282~e^{i 1.862\pi}$ & $0.554~e^{i 0.948\pi}$\\
$\beta_{1}/\alpha_{1}$ & $0.082~e^{i 0.801\pi}$ & $3.321~e^{i 0.085\pi}$\\
$\beta_{2}/\alpha_{1}$ & --- & $0.014~e^{i 1.995\pi}$\\
$\gamma_{1}/\alpha_{1}$ & $0.299~e^{i 1.180\pi}$ & $0.340~e^{i 0.511\pi}$\\
$\gamma_{2}/\alpha_{1}$ & $0.247~e^{i 0.129\pi}$ & $0.154~e^{i 1.539\pi}$\\
$\gamma_{3}/\alpha_{1}$ & $0.247~e^{i 0.120\pi}$ & $0.149~e^{i 1.546\pi}$\\
$\alpha_1^2 v_u^2/v_R({\rm meV})$ & $8.242$ & $12.535$\\
$\alpha_1 v_u/{\rm GeV}$ & $24.249$ & $15.606$\\
$\alpha_1 r_1 v_d$/{\rm GeV} & $0.313$ & $0.272$\\
\hline\hline
$\sin^2\theta_{12}^l$ & $0.313$ & $0.309$\\
$\sin^2\theta_{13}^l$ & $0.02241$ & $0.02247$\\
$\sin^2\theta_{23}^l$ & $0.443$ & $0.430$\\
$\delta_{CP}^l/^\circ$ & $280.211$ & $318.757$\\
$\alpha_{21}/^\circ$ & $199.009$ & $163.502$\\
$\alpha_{31}/^\circ$ & $237.526$ & $205.183$\\
\hline $m_e/m_\mu$ & $0.00467$ & $0.00471$\\
$m_\mu/m_\tau$ & $0.0592$ & $0.0571$\\
$m_1/{\rm meV}$ & $2.009$ & $3.206$\\
$m_2/{\rm meV}$ & $8.845$ & $9.191$\\
$m_3/{\rm meV}$ & $50.036$ & $50.132$\\
$m_{\beta\beta}/{\rm meV}$ & $0.374$ & $0.325$\\
$M_1/{\rm GeV}$ & $2.308\times 10^{11}$ & $8.367\times 10^9$\\
$M_2/{\rm GeV}$ & $5.582\times 10^{12}$ & $1.934\times 10^{12}$\\
$M_3/{\rm GeV}$ & $1.023\times 10^{14}$ & $1.641\times 10^{13}$\\
$v_R/{\rm GeV}$ & $7.134\times 10^{13}$ & $1.943\times 10^{13}$\\
\hline $\theta_{12}^q$ & $0.229$ & $0.229$\\
$\theta_{13}^q$ & $0.00535$ & $0.00393$\\
$\theta_{23}^q$ & $0.0414$ & $0.0366$\\
$\delta_{CP}^q/^\circ$ & $86.767$ & $65.647$\\
\hline $m_u/m_c$ & $0.00269$ & $0.00250$\\
$m_c/m_t$ & $0.00268$ & $0.00292$\\
$m_d/m_s$ & $0.0306$ & $0.0601$\\
$m_s/m_b$ & $0.0171$ & $0.0177$\\
\hline $m_b/m_\tau$ & $0.767$ & $0.672$\\
\hline $Y^{\rm num}_{B}$ & $8.681\times 10^{-11}$ & $8.454\times 10^{-11}$\\
$Y_B^{\rm ap}$ & $-6.378\times 10^{-11}$ & $-1.179\times 10^{-12}$ \\
\hline $\chi^2_l$ & $3.151$ & $8.969$\\
$\chi^2_q$ & $41.647$ & $16.306$\\
$\chi^2_{b\tau}$ & $1.540$ & $3.775$\\
$\chi^{2}_{Y_{B}}$ & $0.021$ & $0.971$ \\
$\chi^2_{total}$ & $46.359$ & $30.021$\\
\hline \hline
\end{tabular}
\caption{\label{tab:fit-non_minimal_models-fail}
Benchmark points at which the analytical approximation of section~\ref{subsec:analytical-app} fails for the concerned models. The vanishing initial abundance of right-handed neutrinos is assumed in \texttt{BM1} model while thermal initial abundance is assumed in \texttt{BM2}. }
\end{table}

\end{appendix}

\clearpage

\providecommand{\href}[2]{#2}\begingroup\raggedright\endgroup

\clearpage

\begin{figure}[hptb!]
\centering
\includegraphics[width=6.5in]{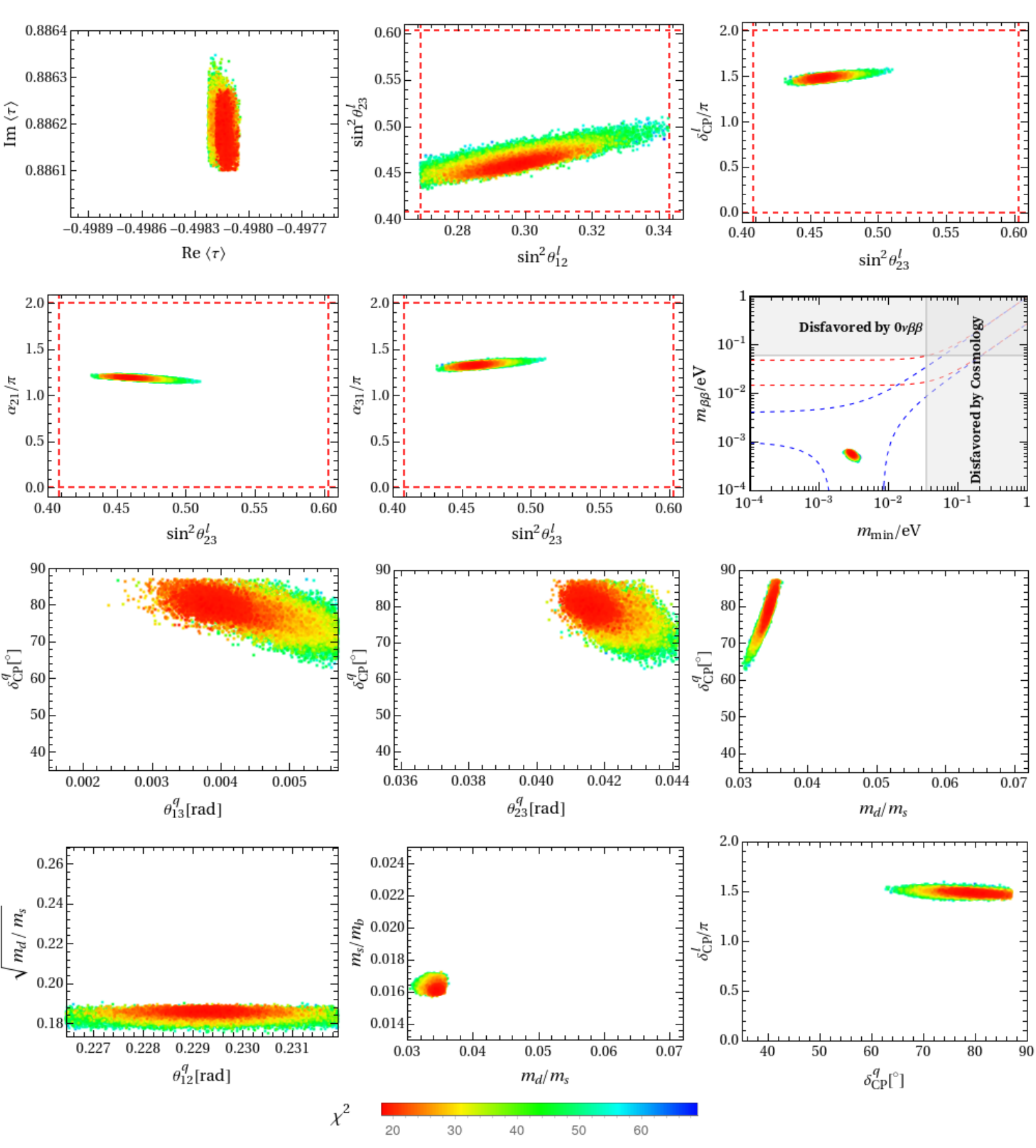}
\caption{The values of the complex modulus $\tau$ compatible with experimental data and the correlations between the neutrino mixing angles, CP violation phases, quark mass ratios and mixing parameters for the model \texttt{BM1} with vanishing initial condition. The vertical and horizontal dashed lines are the $3\sigma$ bounds taken from~\cite{Esteban:2020cvm}. Note that the contribution of $Y_B$ to  $\chi^2$ is not included here. }
\label{fig:non_minimal_2-VIA}
\end{figure}

\begin{figure}[hptb!]
\centering
\includegraphics[width=0.95\textwidth]{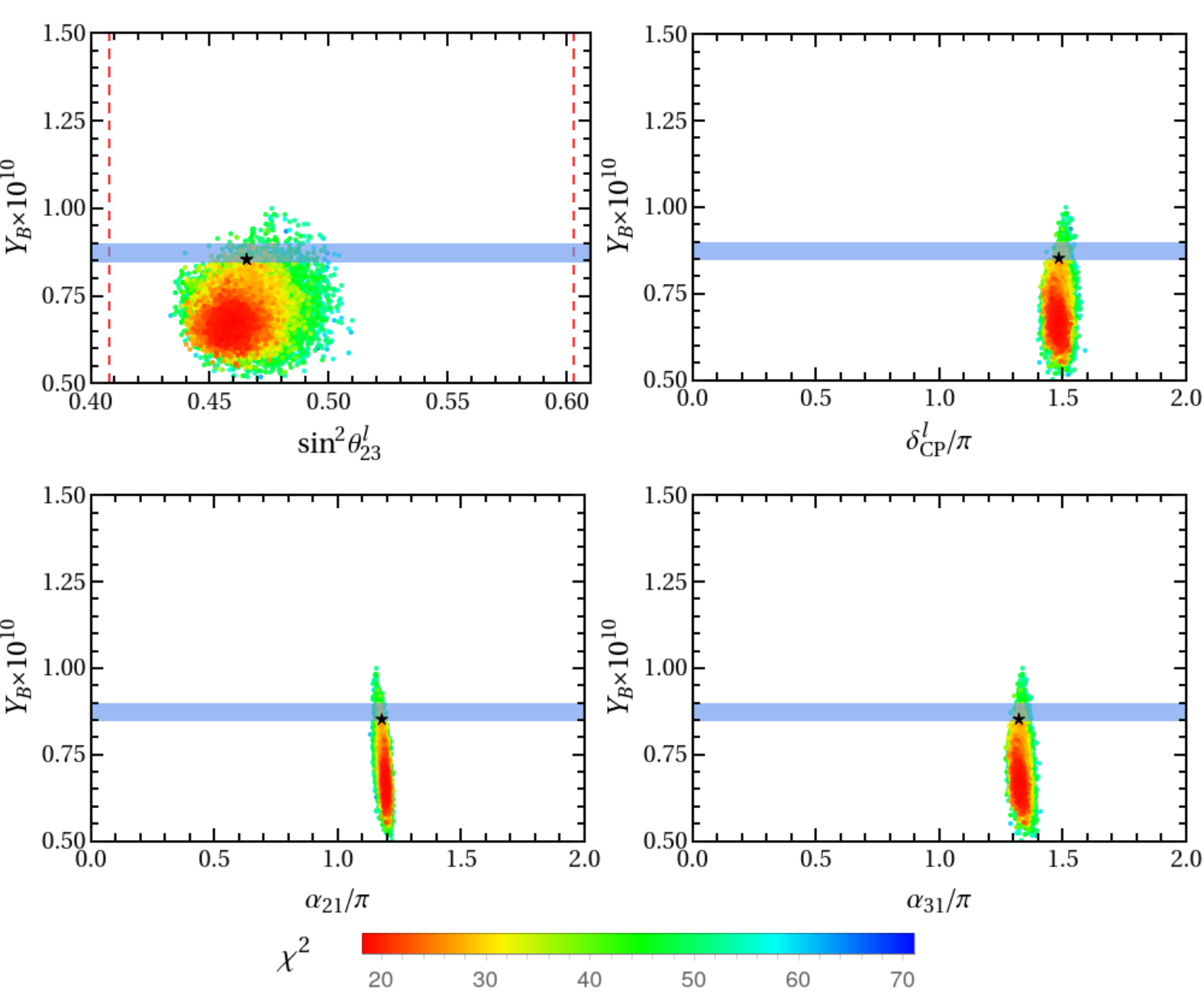}
\caption{The numerical results for the correlation between the baryon asymmetry $Y_B$ and $\sin^2\theta^l_{23}$, $\delta^{l}_{CP}$, $\alpha_{21}$, $\alpha_{31}$ in the \texttt{BM1} model with vanishing initial condition. The horizontal band denotes the experimentally allowed region of $Y_B$. The star ``{$\bigstar$}''  stands for the best fitting point in table~\ref{tab:fit-non_minimal_model_2-VIA}. Note that the contribution of $Y_B$ to  $\chi^2$ is not included here.  }
	\label{fig:non_minimal_2-YB-VIA}
\end{figure}

\begin{figure}[hptb!]
\centering
\includegraphics[width=6.5in]{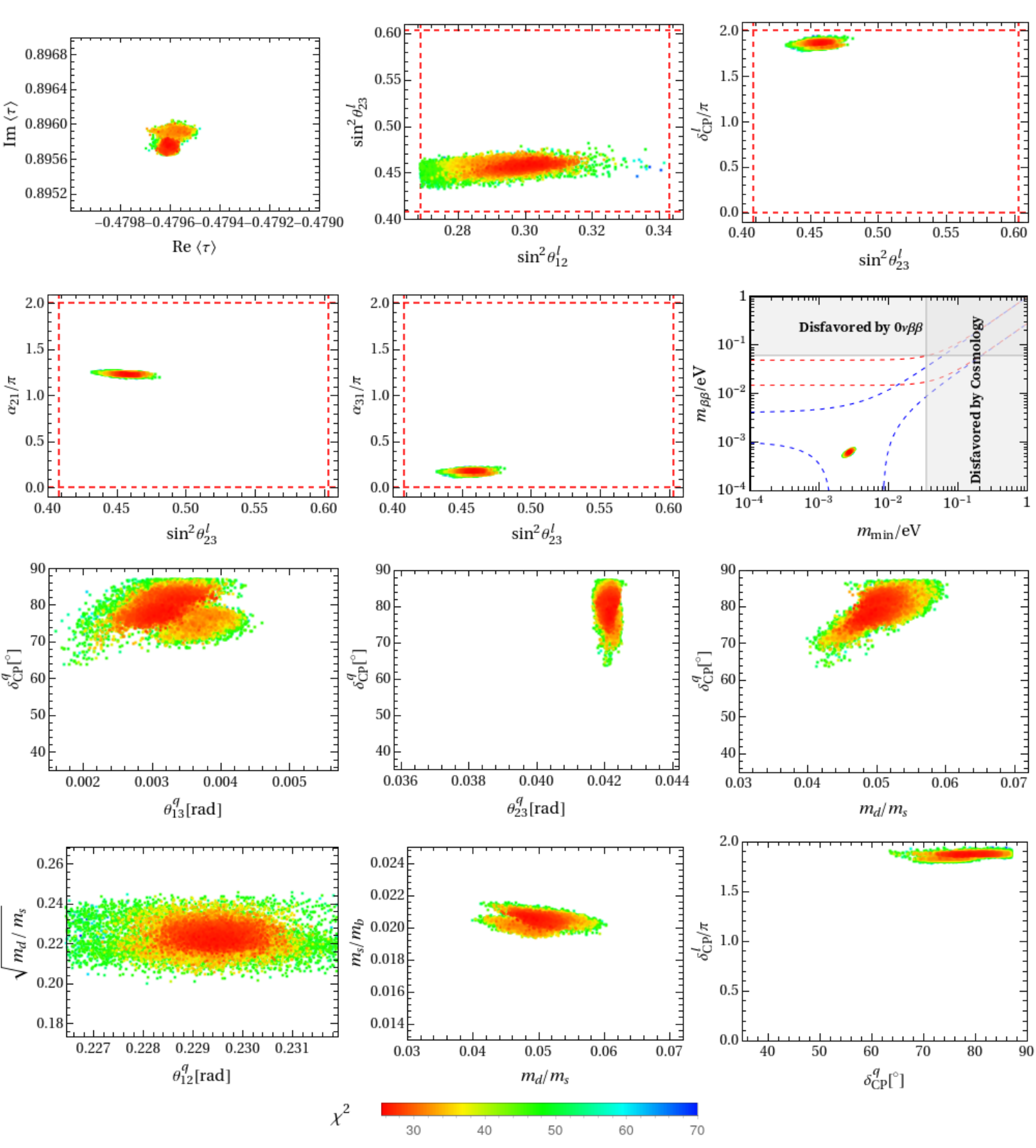}
\caption{The values of the complex modulus $\tau$ compatible with experimental data and the correlations between the neutrino mixing angles, CP violation phases, quark mass ratios and mixing parameters for the model \texttt{BM2} with vanishing initial condition. The vertical and horizontal dashed lines are the $3\sigma$ bounds taken from~\cite{Esteban:2020cvm}. Note that the contribution of $Y_B$ to  $\chi^2$ is not included here. }
\label{fig:non_minimal_3-VIA}
\end{figure}

\begin{figure}[hptb!]
\centering
\includegraphics[width=6.5in]{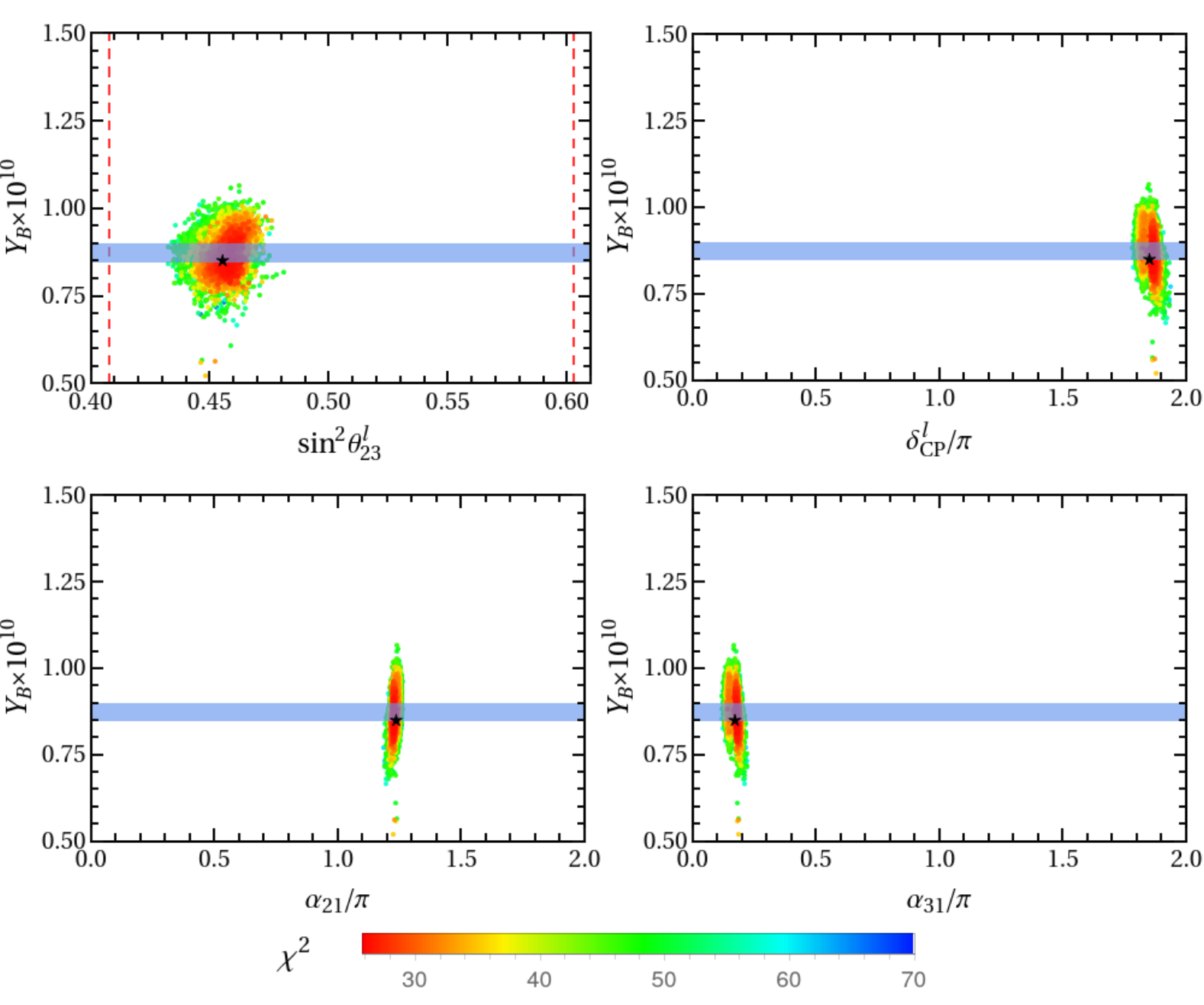}
\caption{The numerical results for the correlation between the baryon asymmetry $Y_B$ and $\sin^2\theta^l_{23}$, $\delta^{l}_{CP}$, $\alpha_{21}$, $\alpha_{31}$ in the \texttt{BM2} model with vanishing initial condition. The horizontal band denotes the experimentally allowed region of $Y_B$. The star ``{$\bigstar$}''  stands for the best fitting point in table~\ref{tab:fit-non_minimal_model_2-VIA}. Note that the contribution of $Y_B$ to  $\chi^2$ is not included here.  }
	\label{fig:non_minimal_3-YB-VIA}
\end{figure}

\begin{figure}[hptb!]
\centering
\includegraphics[width=6.5in]{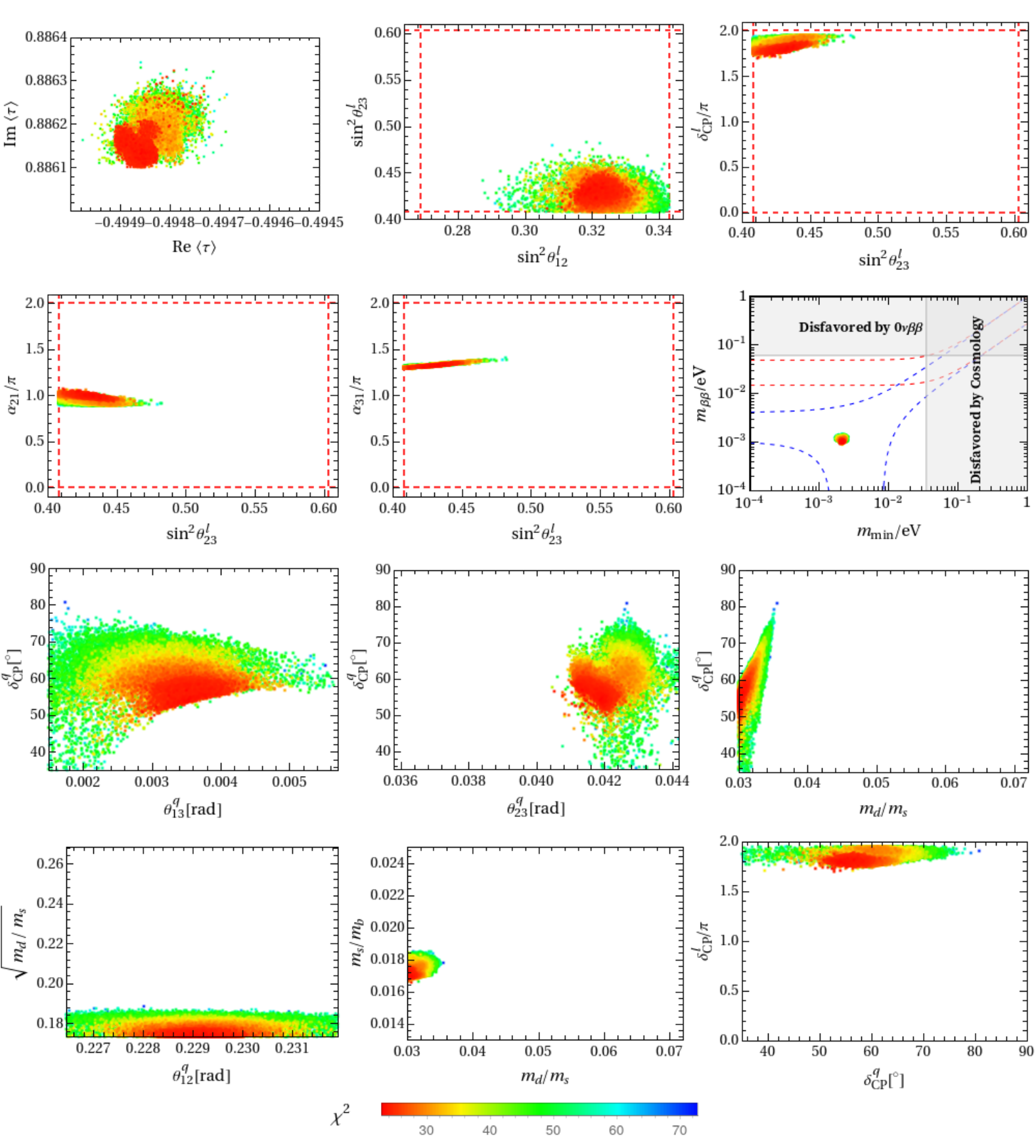}
\caption{The values of the complex modulus $\tau$ compatible with experimental data and the correlations between the neutrino mixing angles, CP violation phases, quark mass ratios and mixing parameters for the model \texttt{BM1} with thermal initial condition. The vertical and horizontal dashed lines are the $3\sigma$ bounds taken from~\cite{Esteban:2020cvm}. Note that the contribution of $Y_B$ to  $\chi^2$ is not included here.  }
\label{fig:non_minimal_2-TIA}
\end{figure}

\begin{figure}[hptb!]
\centering
\includegraphics[width=5.5in]{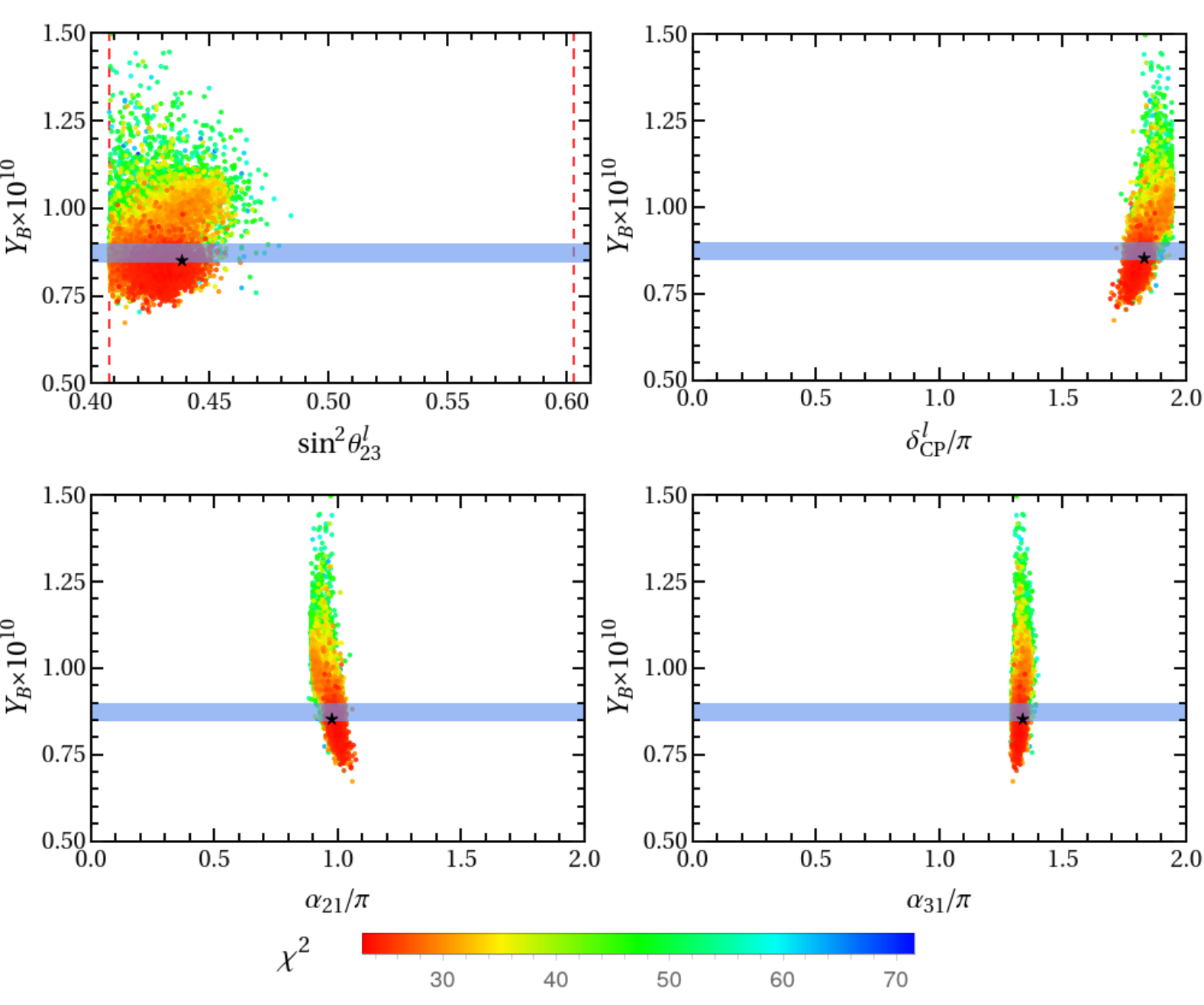}
\caption{The numerical results for the correlation between the baryon asymmetry $Y_B$ and $\sin^2\theta^l_{23}$, $\delta^{l}_{CP}$, $\alpha_{21}$, $\alpha_{31}$ in the \texttt{BM1} model with thermal initial condition. The horizontal band denotes the experimentally allowed region of $Y_B$. The star ``{$\bigstar$}''  stands for the best fitting point in table~\ref{tab:fit-non_minimal_model_2_second-TIA}. Note that the contribution of $Y_B$ to  $\chi^2$ is not included here.  }
\label{fig:non_minimal_2-YB-TIA}
\end{figure}

\begin{figure}[hptb!]
\centering
\includegraphics[width=6.5in]{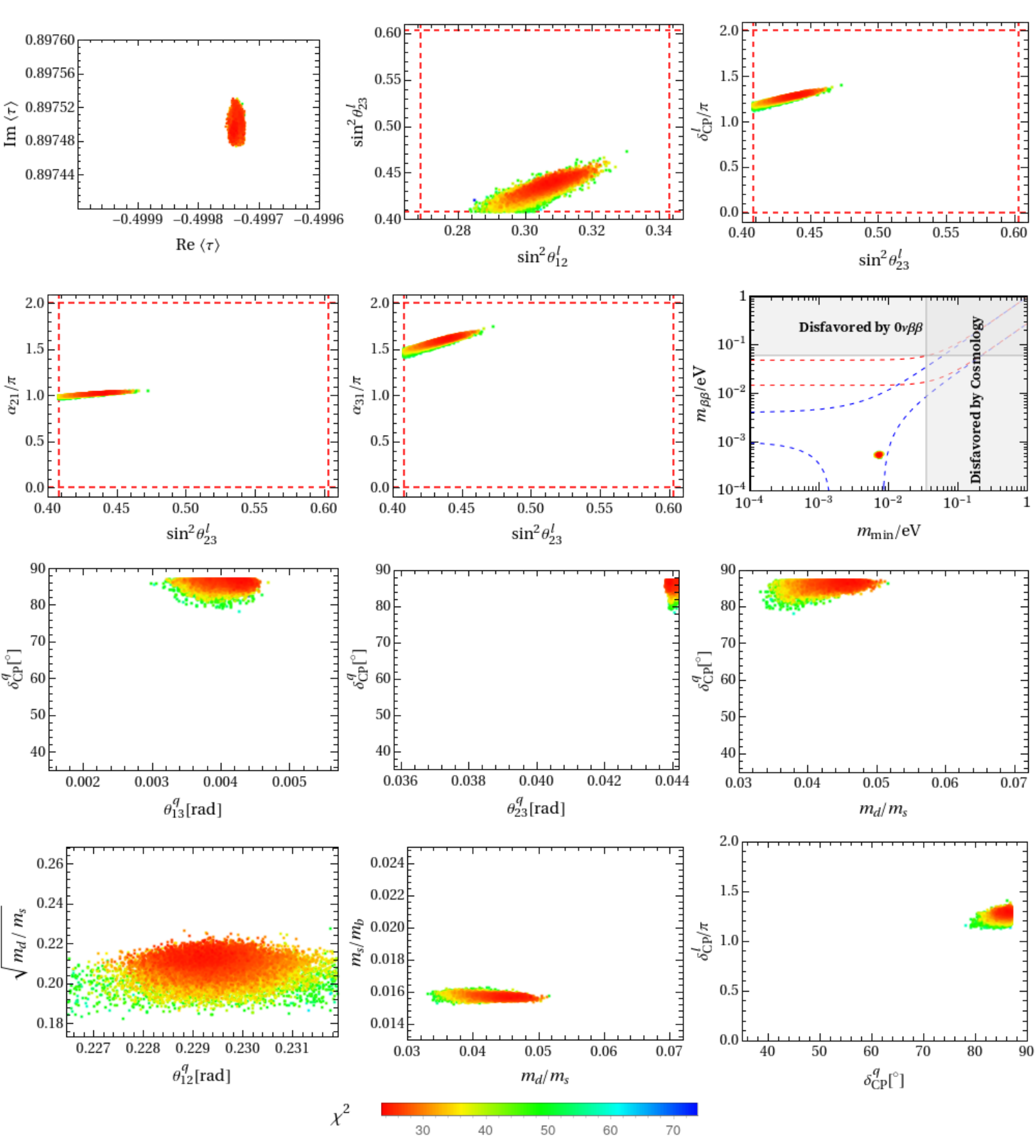}
\caption{The values of the complex modulus $\tau$ compatible with experimental data and the correlations between the neutrino mixing angles, CP violation phases, quark mass ratios and mixing parameters for the model \texttt{BM2} with thermal initial condition. The vertical and horizontal dashed lines are the $3\sigma$ bounds taken from~\cite{Esteban:2020cvm}. Note that the contribution of $Y_B$ to  $\chi^2$ is not included here.  }
\label{fig:non_minimal_3-TIA}
\end{figure}

\begin{figure}[hptb!]
\centering
\includegraphics[width=5.5in]{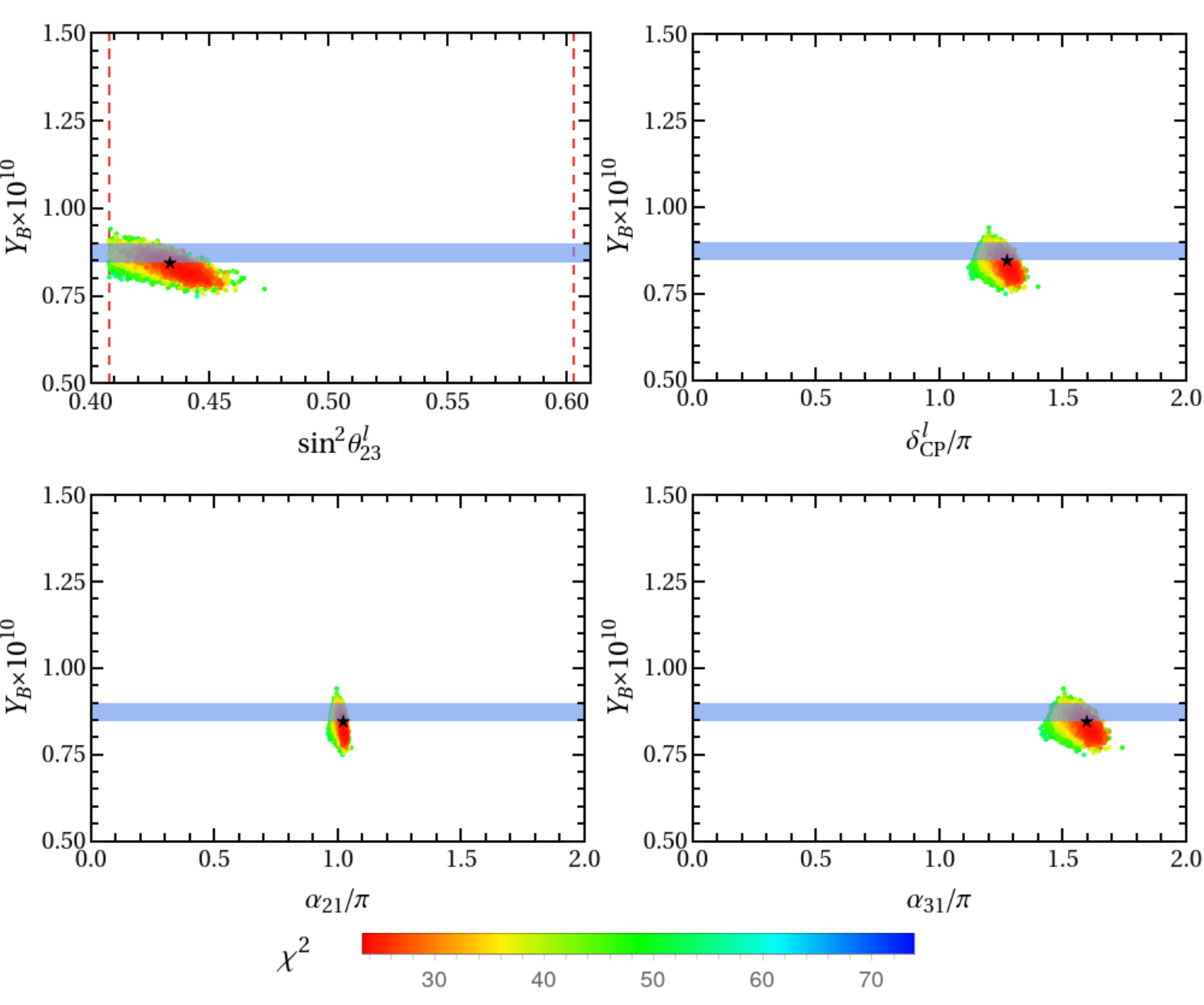}
\caption{The numerical results for the correlation between the baryon asymmetry $Y_B$ and $\sin^2\theta^l_{23}$, $\delta^{l}_{CP}$, $\alpha_{21}$, $\alpha_{31}$ in the \texttt{BM2} model with thermal initial condition. The horizontal band denotes the experimentally allowed region of $Y_B$. The star ``{$\bigstar$}''  stands for the best fitting point in table~\ref{tab:fit-non_minimal_model_2_second-TIA}. Note that the contribution of $Y_B$ to  $\chi^2$ is not included here. }
\label{fig:non_minimal_3-YB-TIA}
\end{figure}

\end{document}